\documentclass[preprint,authoryear,12pt]{rarticle}

\usepackage{graphics}
\usepackage{epsfig}
\usepackage{psfrag}
\usepackage{bm}
\usepackage{subfigure}

\usepackage{amssymb}
 \usepackage{amsthm}
\usepackage{amsmath}

\usepackage{color}

\begin{document}

\begin{frontmatter}



\title{Predicting  the  pressure-volume curve of an elastic  microsphere composite}


\author{ Riccardo De Pascalis, I.\ David Abrahams, William J.\  Parnell}

\address{School of Mathematics, University of Manchester,\\ Manchester M13 9PL,  United Kingdom\\}

\begin{abstract}
The effective macroscopic response of nonlinear elastomeric inhomogeneous materials is of great interest in many applications including nonlinear composite materials and soft biological tissues. The interest of the present work  is associated with a \textit{microsphere composite material}, which is modelled as a matrix-inclusion composite. The matrix phase is a homogeneous isotropic nonlinear rubber-like material and the inclusion phase is more complex, consisting of a distribution of sizes of stiff thin spherical shells filled with gas. Experimentally, such materials have been shown to undergo complex deformation under cyclic loading.  Here, we consider microspheres embedded in an unbounded host material and assume that a hydrostatic pressure is applied in the `far-field'.  Taking into account a variety of effects  including buckling of the spherical shells, large deformation of the host phase and evolving microstructure, we derive a model predicting the  \textit{pressure-relative volume change} load curves. Nonlinear constitutive behaviour of the matrix medium is accounted for by employing neo-Hookean and Mooney-Rivlin incompressible models.  Moreover a nearly-incompressible  solution is derived via asymptotic analysis for a spherical cavity embedded in un unbounded isotropic homogeneous hyperelastic medium loaded hydrostatically. The load-curve predictions reveal a strong dependence on the microstructure of the composite, including distribution of microspheres,  the stiffness of the shells, and on the initial volume fraction of the inclusions, whereas there is only a modest dependence on the characteristic properties of the nonlinear elastic model used for the rubber host.
\end{abstract}


\end{frontmatter}


\section{Introduction}
\label{introduction}
\begin{figure}[ht]
	\centering
	\subfigure[]{
	    \includegraphics[scale=0.26]{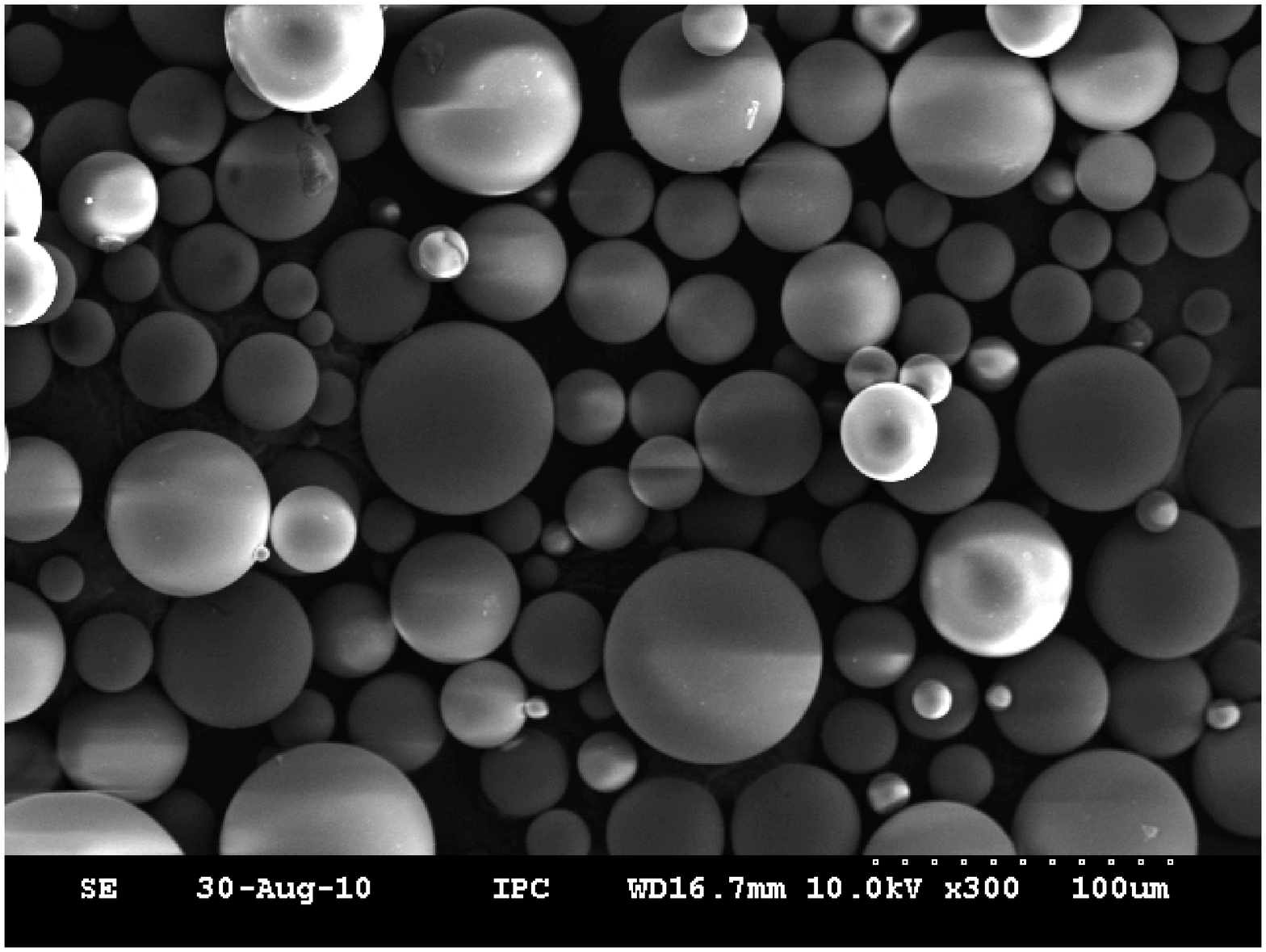}
	    \label{figures/14.eps}
	}
	\subfigure[]{
        \includegraphics[scale=0.26]{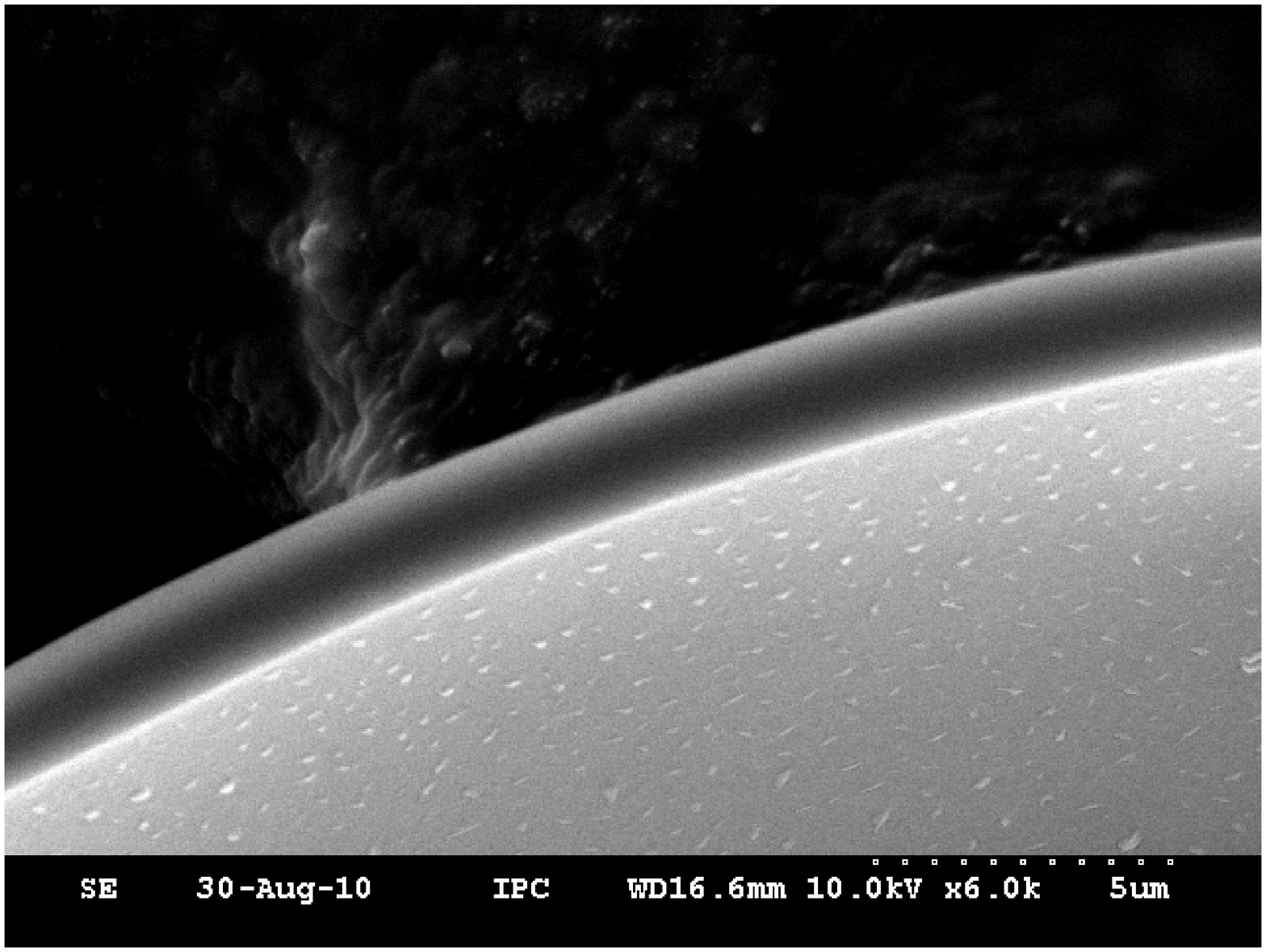}
	    \label{figures/15.eps}
	}
	\caption[]{Scanning electron microscopy image of (a) a Hollow Glass Microsphere composite material and (b) its surface/shell structure. In this situation (with applications to thermal conductivity) the composite is densely filled with microspheres. Reproduced (with kind permission) from \cite{li11}.}
	\label{fig:HGM}
	\end{figure}

Microsphere composites are used in a multitude of industrial applications. Good examples are ultra-low density fillers in engineering materials such as composites, coatings, sealants, explosives, automotive components, paint and crack fillers and elastomers and as blowing agents in printing inks \citep{ash07}. The microsphere is a spherical particle, a few microns in diameter, with a thermoplastic shell, and shell to diameter ratio typically of the order of $0.01$. The use of microspheres in materials brings forth numerous benefits which include reducing density, 
improving stability, increasing impact strength, providing a smoother surface finish, increasing thermal insulation, increasing compressibility and often reducing costs. A scanning electron microscopy image of the microstructure of a hollow glass microsphere composite (with very high volume fraction) is shown in Fig.\ \ref{fig:HGM} taken from \cite{li11}.

\begin{figure}[ht]
	\centering
	\subfigure[]{
	    \includegraphics[scale=0.55]{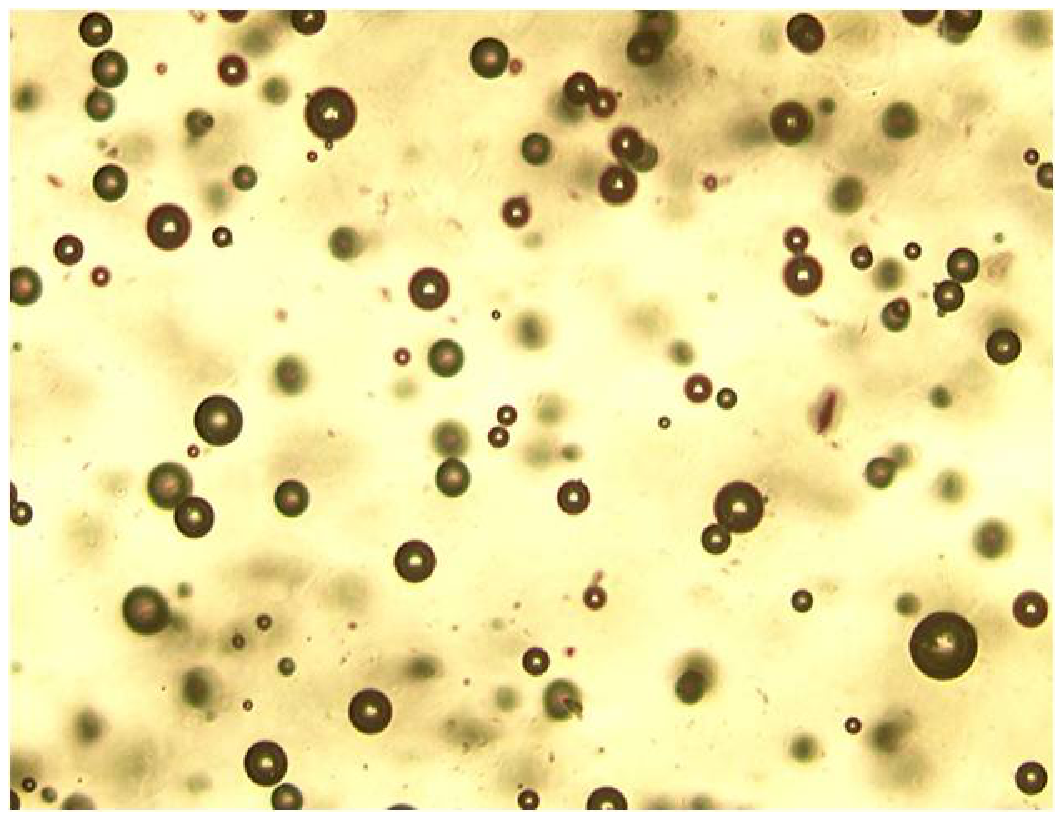}
	    \label{figures/14p.eps}
	}
	\subfigure[]{
\psfrag{s}{stress (MPa)}
\psfrag{e}{strain $\%$}
        \includegraphics[scale=0.45]{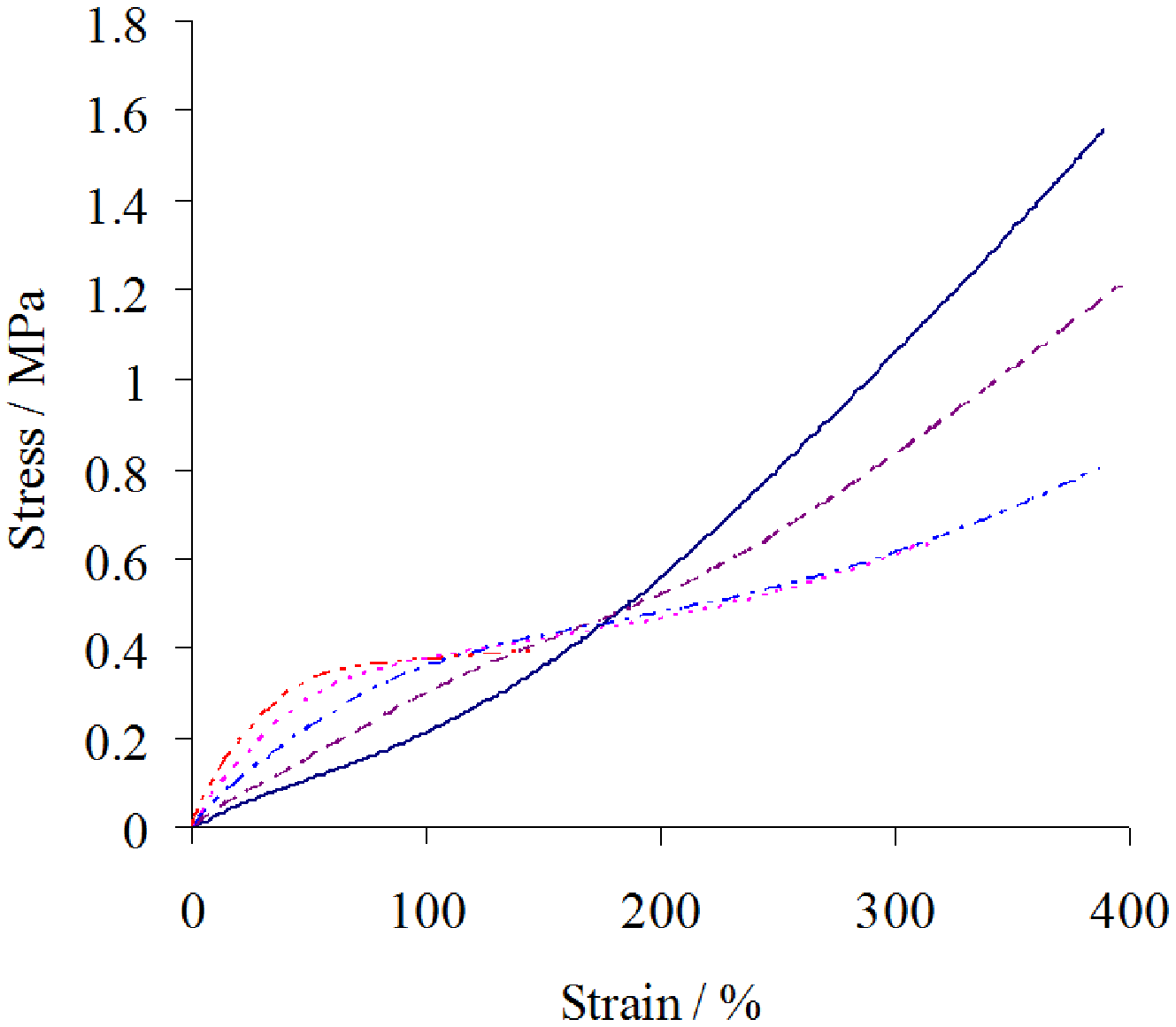}
	    \label{figures/15p.eps}
	}
	\caption[]{In (a) we show an image of a Silicone Room Temperature Vulcanizing (RTV) microsphere elastomer filled with 5\% volume of Expancel microspheres. In (b) we show an experimentally determined stress-strain curve associated with this composite under uniaxial tension. The solid curve is unfilled and others refer to increasing
volume fractions of the microsphere phase.  Reproduced (with kind permission) from \cite{shorter2008}.}
	\label{shorterimage}
	\end{figure}

The application of interest in this paper is that of acoustics, using microsphere composites as a means of reducing sound reflection. The composite consists of an elastomeric matrix phase, inside which are located a large number of randomly distributed Expancel microspheres, see Fig.\ \ref{figures/14p.eps} \cite{shorter2008}.  Of specific interest is how sound reflection can be affected by a macroscopic hydrostatic pressure applied to the microsphere composite. These materials have been found to be useful in such conditions because the presence of the reinforcing shell delays the onset of the cavity collapse and the consequent degradation in the acoustic performance of the composite. In order to understand exactly how the acoustic characteristics of the material are affected by applied pressure, it is necessary to develop models that describe how the composite deforms mechanically under this applied loading. Experimentally it is known that the constitutive \textit{pressure-relative volume change} curve is nonlinear (and hysteretic during unloading) but the dominant physical mechanisms contributing to this nonlinearity are still not fully understood. In Fig.\ \ref{figures/15p.eps} we illustrate the constitutive behaviour of the material with some experimentally determined load curves associated with the composite for increasing volume fractions of the microsphere material \cite{shorter2008}.

A wide range of work on the modelling of microsphere filled composites has appeared in the literature. For the prediction of their acoustic properties \textit{without applied pressure}, a number of models have been proposed, see e.g.\  \cite{Gau82} and \cite{Baird-Kerr-Townend99} in the elastic and viscoelastic case respectively. Such predictions typically rely upon the use of classical multiple scattering models, the most commonly used being those of \cite{Wat-61, Kus-74a, Bos-74, Gau82, Ans-89}. We note that a useful comparison of experiments and various theories can be found in \cite{Ans-93}.

\cite{Gau84} considered the pressure dependence of dynamic moduli, albeit in a simplified case of porous solids, i.e.\ in the absence of the shell phase and thus the effect of the pressure is to reduce the pore size, the main interest lying in the dynamic material properties. Here, although we are certainly interested in the dynamic response, we wish first to understand the origins of nonlinearity in the \textit{pressure-relative volume change} loading curve associated with the composite. Significant work has been carried out on the deformation of porous materials, see e.g.\ \cite{Mac-50} for an early model for associated effective linear elastic properties, and for rubber foam materials, see e.g.\ \cite{Gen59}, \cite{Gib82} and \cite{Lak93}, where the principal mechanisms of deformation are well understood. However, the composite under consideration here has a more complex microstructure due principally to the presence of microspheres and the lack of understanding as to how they behave within the elastomeric substrate under applied pressure. Microspheres are also present in the context of syntactic foams (a material comprising a polymeric matrix filled with microspheres). Such materials open up the possibility of low density materials with high tolerance to damage. Much of the recent modelling work in this area however has focused on the effective \textit{linear} elastic properties of the composite. A variety of static homogenization techniques have been used, see e.g.\ \cite{Bardella-Genna01, Gupta-Woldesenbet04, porfiri-gupta09, Tagliavia-Porfiri-Gupta09}.

Few models deal with the nonlinear response of a microsphere composite under loading. \cite{Ker02} proposed an elasto-plastic model for the load curve and subsequent prediction of dynamic material properties. One criticism of this model would be that plasticity yields permanent deformation. However, it is well acknowledged that although the load-unload curve is hysteretic, when all load has been removed the material (eventually) returns to its original configuration \citep{private}. Therefore it does not appear that plastic deformation is the cause of nonlinearity. In a related application \cite{Pan08} considered the acoustic response of inhomogeneous media under applied pressure, although it appears that the microstructure is rather different from that considered here. In \cite{shorter2008}, \cite{shorter2010} the problem of the buckling of a single, isolated spherical shell was considered (as a result of axial compression, rather than hydrostatic pressure) using Finite Element Analysis and results were subsequently compared with experiments involving a table tennis ball embedded inside a transparent elastomer. Comparisons were made between perfectly bonded and unbonded spherical shells and subsequent buckling response. The principal argument of the paper was to propose that microsphere buckling is a dominant contributor to the nonlinear behaviour of the pressure-relative volume change curve.

The computational work carried out in \cite{shorter2008} suggests that a model incorporating microsphere buckling could successfully predict the nonlinearity of the load curve.
Therefore here we develop a fundamental mathematical model for the loading portion of the pressure-relative volume change curve by incorporating volumetric changes and local microsphere shell buckling effects. We assume that there is a distribution of shell to radius ratio thicknesses and we will suppose that the microspheres are distributed dilutely, so that interaction effects between microspheres can be neglected. Interaction effects will be considered in future work.

In order to incorporate the effect of buckling of the microsphere shell, we must understand how a spherical shell buckles inside an elastic medium under far-field hydrostatic pressure. Although a great deal of classical work exists regarding the buckling of spherical shells (where the imposed pressure is on the surface of the shell itself), see for example  \cite{wesolowski67,Koiter69,wang_ertepinar72} and more recently \cite{Fu98,goriely_benamar05'}, there is a surprising lack of work regarding the buckling of shells (of any geometry) that are embedded inside another medium. Of specific interest is how the host medium affects the classical buckling pressure.

Initial work into the buckling pressure of a spherical shell embedded in an unbounded uniform elastic medium
 has been carried out by \cite{fox-allwright01} and \cite{Jones-Chapman-Allwright07}. We shall discuss these models and their assumptions later on in the paper, particularly that of \cite{fox-allwright01} which is the model that we shall adopt for buckling here. As described above, \cite{shorter2008} also carried out some experimental work related to this problem.



The fundamental objective here then is to determine a model for the loading portion of the \textit{pressure-relative volume change curve} by considering a distribution of shell thickness to radius ratio of microspheres which are dilutely dispersed throughout the material. We also introduce nonlinear (finite) elasticity in order to incorporate large deformation of the rubber composite in the post-buckling regime.
The theory provides a modelling tool to assess certain likelihood scenarios. In particular we are able to assess the sensitivity of effective properties to changes in specific parameters, e.g.\ distribution of microspheres, nonlinearity and constitutive behaviour of the constituent materials that make up the composite and the gas law inside microspheres.

\section{Preliminaries and background}
\label{Preliminaries and background}

We consider a composite material with two constituents (or \textit{phases} as we shall term them here) known as the matrix and inclusion phase. The matrix phase is a (possibly compressible) homogeneous rubber material and the inclusion phase consists of a distribution of thin spherical shells (possibly filled with some gas) of initial radius $A$ and shell thickness $H$. We allow for the possibility of a distribution of microsphere shell thickness to radius ratios $X=H/A$. This distribution is governed by a probability distribution function $F(X)$. The volume fraction of the inclusion phase is denoted by $\Phi$. We are specifically interested in the problem where the material is subjected to an external hydrostatic pressure $\hat{p}$. We shall state all pressures relative to atmospheric pressure $p_{\textsc{{atm}}}$ so that upon defining $p=\hat{p}-p_{\textsc{{atm}}}$, $p=0$ corresponds physically to atmospheric pressure. Similarly, we assume that the gas inside the microspheres is also initially at atmospheric pressure so that denoting $\hat{p}_{\textsc{{in}}}$ as the internal hydrostatic pressure we can initially set $p_{\textsc{{in}}}=\hat{p}_{\textsc{{in}}}-p_{\textsc{{atm}}}=0$. Henceforth all pressures are thus defined relative to atmospheric pressure.

\begin{figure}
\centering
\psfrag{p}{\scriptsize{$p$}}
\psfrag{d}{\scriptsize{deformation}}
\includegraphics[scale=0.4]{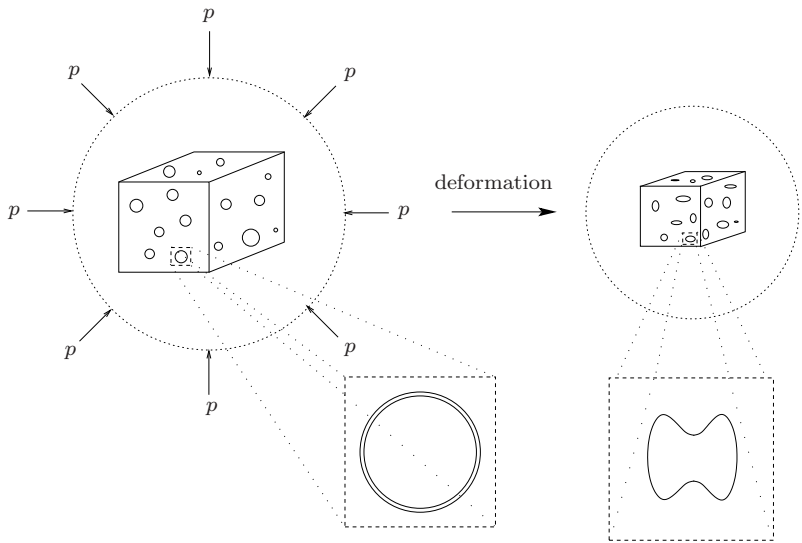}
\caption{A cube of the microsphere composite material (no scale is implied) is subjected to hydrostatic pressure $p$ in the far field with inset figure indicating that the microspheres are spherical pre-buckling. This spherical symmetry is retained until the onset of buckling, after which the shells deform significantly, losing their rigidity in this state, as indicated in the figure inset. We do not indicate the precise structure of the shell post-buckling: this requires detailed post-buckling analysis.}
\label{sample_cube}
\end{figure}

Consider for the moment a single microsphere embedded in an \textit{unbounded} matrix material so that we assume that the pressure is applied in the `far-field'. At a critical far-field pressure $p_c$, this shell will buckle and the microsphere will lose its compressive rigidity for $p>p_c$. Since we assume that we have a distribution of microspheres, each with a different $X=H/A$, the microspheres will buckle successively as the external pressure is continuously increased. We illustrate this in Fig.\ \ref{sample_cube}. The prediction of the pressure-volume relation, given the volume fraction $\Phi$ of microspheres, the constitutive behaviour of the matrix phase,  the elastic properties of the microsphere shell, the knowledge of gas internal to the microspheres and the overall distribution of $X$, is clearly a non-trivial problem. The effects of interaction on buckling have thus far not been studied and therefore here we consider the case where buckling of a microsphere depends only on the far-field hydrostatic pressure $p$ and \textit{not} on the influence of other microspheres. We anticipate therefore that this model is valid for a dilute dispersion of microspheres. We note however, that in many homogenization theories, it is often surprising how accurate dilute dispersion theories are even in the non-dilute regime \citep{Parnell-Abrahams-Brazier-Smith10}. Later work will consider interaction effects in more detail.

\begin{figure}
\centering
\psfrag{S}{\scriptsize{$S$}}
\psfrag{D}{\scriptsize{$A-H$}}
\psfrag{A}{\scriptsize{$A$}}
\psfrag{p}{\scriptsize{$p$}}
\includegraphics[scale=0.5]{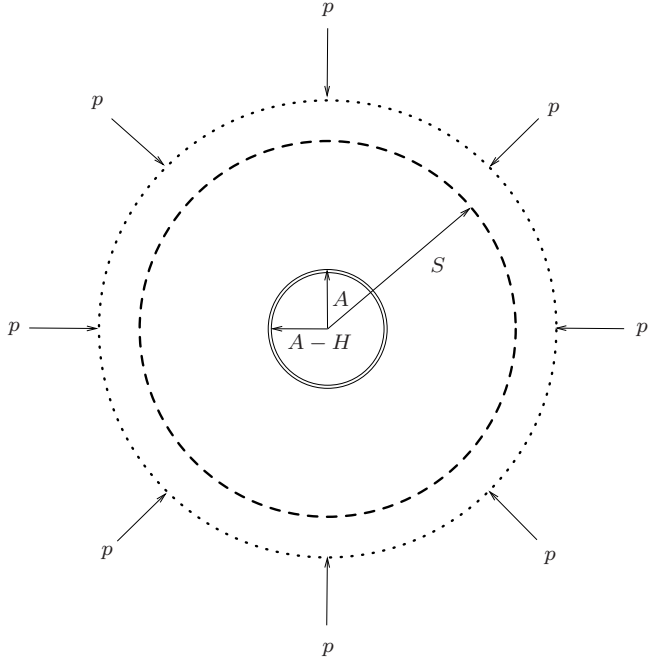}
\caption{A single `composite sphere' loaded by the hydrostatic far-field pressure $p$.}
\label{singleCS}
\end{figure}

We shall consider each microsphere to have a fixed initial radius $A$ and let the shell thickness $H$ vary, so that $X=H/A$ is governed by a probability distribution function $F(X)$. Alternatively we could consider $H$ fixed and vary $A$ but it transpires that the analysis of the former is more straightforward. (Note that for a single microsphere inclusion the scale invariance means that varying $A$ or $H$ for any fixed $X$ must yield the same result.) In some cases we need to refer to the middle surface of the shell whose radius we denote by $\hat{A}=A-H/2$, and  the shell thickness to middle radius ratio as $\hat{X}=H/\hat{A}$. Also  the probability distribution function can therefore be given in terms of $\hat{X}$, i.e.\ $F(\hat{X})$.  With reference to Fig.\ \ref{singleCS}  we define a (fictitious) radius $S>A$ by the condition $\Phi=(A/S)^3$ where $\Phi$ is the prescribed volume fraction of microspheres. We then consider how the material deforms, and how the microsphere buckles, given some hydrostatic pressure $p$ in the far-field with this region inside $S$ (containing the microsphere and which we will term the \textit{composite sphere} (CS)) embedded in a purely matrix material. The prescription of the radius $S$ allows us to consider the volume change of the matrix region under compression.

We denote by $\kappa,\mu,E,\nu$ the bulk, shear and Young's moduli and Poisson's ratio respectively from linear elasticity and we note the relations $\nu=(3\kappa-2\mu)/(2(3\kappa+\mu))$ and $E=9\kappa\mu/(3\kappa+\mu)$ since later we usually specify $\mu$ and $\kappa$. We will make use of the subscripts $s$ and $m$  when we wish to refer to the shell and to the matrix medium, respectively.

Before the microsphere shell buckles (which we shall term the \textit{pre-buckling} stage) we consider the elastic behaviour of both the matrix and microsphere shell to be linear. As will be shown, this is reasonable since the shell stiffness is significantly higher than that of the matrix phase and therefore induced strains in both media will be small (see also section 7 in \cite{Jones-Chapman-Allwright07} for more details). After the microsphere shell buckles (which we term the \textit{post-buckling} stage) we make the assumption that the shell will lose almost all of its rigidity and therefore that the post-buckled microsphere can be replaced by a cavity (whilst still ensuring continuity of displacement and traction between the matrix and fluid as we shall show later).  In this post-buckling regime we incorporate nonlinear elastic behaviour by permitting large strains and also nonlinear constitutive behaviour.

\begin{figure}
\centering
\psfrag{0.5}[c][l]{\small{$0.5$}}
\psfrag{1.0}[c][l]{\small{$1.0$}}
\psfrag{1.5}[c][l]{\small{$1.5$}}
\psfrag{2.0}[c][l]{\small{$2.0$}}
\psfrag{0.005}[c][b]{\small{$0.005\hspace{0.4cm}$}}
\psfrag{0.010}[c][b]{\small{$0.010\hspace{0.4cm}$}}
\psfrag{0.015}[c][b]{\small{$0.015\hspace{0.4cm}$}}
\psfrag{0.020}[c][b]{\small{$0.020\hspace{0.4cm}$}}
\psfrag{0.025}[c][b]{\small{$0.025\hspace{0.4cm}$}}
\psfrag{0.1}[c][l]{\small{$0.1$}}
\psfrag{0.2}[c][l]{\small{$0.2$}}
\psfrag{0.3}[c][l]{\small{$0.3$}}
\psfrag{0.4}[c][l]{\small{$0.4$}}
\psfrag{B}{\footnotesize{$-\dfrac{u_r^m}{A}$}}
\psfrag{A}{\footnotesize{$\dfrac{p}{\mu_m}$}}
\includegraphics[scale=0.75]{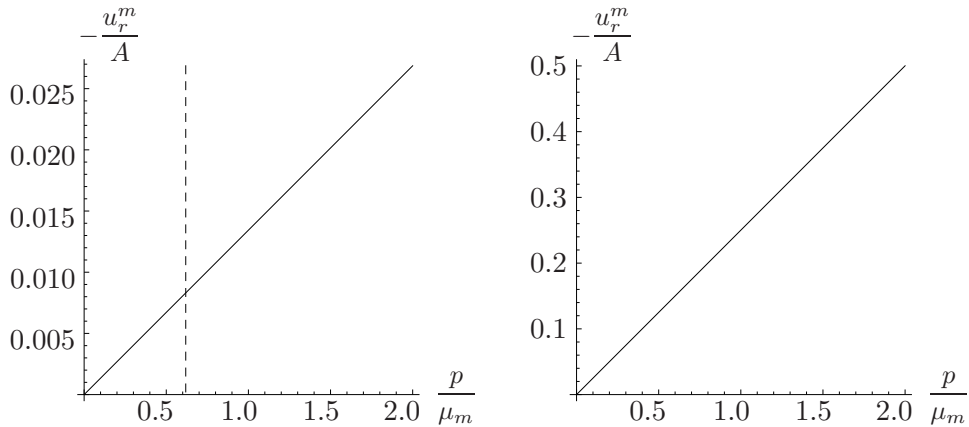}
\caption{Plot of the the scaled radial displacement $u(A)_r^m/A$ determined using linear elasticity when the shell is (left) and is not (right) included in the model. Associated values of $X$ used are $X=0.01$ and $X=0$ respectively. The dashed line is the critical pressure predicted by  \eqref{F-A-criteria}.}
\label{displacement}
\end{figure}

In order to justify the linear pre-buckling and nonlinear post-buckling assumptions, respectively, we consider the following example. We are here interested  in understanding how the stiffness of the shell can make the material more rigid as compared with the case when the shell is absent. To this end  we can consider for example the case of a thin glassy shell (see \cite{Baird-Kerr-Townend99}), with surrounding polymeric elastomer composed of a polyurethane material. The matrix Young's modulus can be taken as $E_m=3.6$ MPa and Poisson ratio is typically close to $0.5$ \citep{Diaconu-Dorohoi05}. In terms of bulk and shear moduli, we choose the parameter set
\begin{align}\label{material-constants}
\mu_s &= 1.26\ \textrm{GPa},& \kappa_s &= 2.1\ \textrm{GPa}, \notag\\
\mu_m &=1.2\  \textrm{MPa},&\kappa_m &= 4\ \textrm{GPa}.
\end{align}
Note that with this choice,  $\nu_m=0.49985$ i.e.\ the matrix is considered essentially incompressible.  Next, let us take a microsphere with shell to radius ratio $X=0.01$ and for an imposed scaled far-field pressure $p/\mu_m$, we evaluate the scaled displacement (with notation reported in section \ref{Linear elasticity}) $u_r^m(A)/A$ at $r=A$ (the radius at which the displacement is maximum). In Fig.\ \ref{displacement} we plot this maximum displacement as a function of the imposed pressure when the shell is and is not present (left and right on the figure, respectively) and in both cases we note that this is predicted by linear elasticity theory. The dashed line denotes the critical pressure for this shell to radius ratio, predicted by the Fok-Allwright buckling criterion \eqref{F-A-criteria} \citep{fox-allwright01}.  When the shell \textit{is} included, values of $u(A)_r^m/A$ remain small for applied pressures below this critical value (see the left side of Fig.\ \ref{displacement}). On the contrary, in the absence of the shell,  when reasonable values of the pressure are applied, the linear theory is no longer appropriate due to the large values of the scaled displacements $u_r^m(A)/A$ in this case (see the right side of Fig.\ \ref{displacement}). We conclude that in this latter regime we \textit{must} therefore incorporate full nonlinearity in order to permit finite deformations.

Although we are concerned with matrix materials that ostensibly behave incompressibly (typically $\mu_m/\kappa_m$ is small, in particular  of the order of  $10^{-4}$, see \cite{ogden76}), in section \ref{Post-buckling with slight compressibility} we also calculate the volume change post-buckling for a \textit{nearly-incompressible} theory. As will be shown, it is difficult to distinguish the difference between results for the slightly-compressible and incompresssible cases, as should be expected.

\section{Pre-buckling behaviour and the buckling model}
\label{Pre-buckling behaviour}
In the pre-buckling stage, we consider both shell and matrix phase to be compressible linear elastic materials (noting that the matrix is almost incompressible) which are perfectly bonded, and we assume that gas resides inside the microsphere providing a constant internal pressure $p_{\textsc{{in}}}$. We wish to determine the total volume change (relative to the initial volume) of the material and in order to do this we determine the volume change in the composite sphere (CS) when the pressure $p$ is imposed at infinity.

\subsection{Linear elasticity}
\label{Linear elasticity}

Under the assumption of linear isotropic elasticity, the governing equations of the corresponding static  boundary value problem with no body forces are given as follows
\begin{equation}\label{linear_equilibrium}
\sigma_{ij,i}=0,
\end{equation}
\begin{equation}\label{strain_tensor}
\epsilon_{ij}=\dfrac{1}{2}\left(u_{i,j}+u_{j,i}\right),
\end{equation}
\begin{equation}\label{linear_constitutive_equations}
\sigma_{ij}=\lambda\epsilon_{kk}\delta_{ij}+2\mu \epsilon_{ij},
\end{equation}
where $\sigma_{ij}, \epsilon_{ij}$,  and $u_i$  are the components of the stress and strain tensors,  and the displacements, respectively and we have introduced the Kronecker delta tensor $\delta_{ij}$. The matrix material is homogeneous with Lam\'e constants $\lambda,\mu$ where $\lambda=\kappa-2\mu/3$. Since the problem is linearly elastic, geometry is spherically symmetric and a purely radial stress is applied, then spherical symmetry is preserved ($u_r=u_r(r), u_{\theta}=u_{\phi}=0$). Hence, equation (\ref{linear_equilibrium}) reduces  to  a single second order ordinary differential equation which is independent of the Lam\'{e} moduli. The general solution for the displacement in the shell and medium region is therefore of the form
\begin{equation}\label{solution}
u_r^{i}(r)=A_i r + \frac{B_i}{r^2},
\end{equation}
where $i=s,m$ refers to the shell and  matrix respectively, and $A_i,B_i$ are constants that are fully determined by
imposing continuity of displacement and radial stress on $r=A$ and the following loading boundary conditions
\begin{equation}
\sigma^s_{rr}(A-H)=-p_{\textsc{{in}}},\quad \sigma^m_{rr}(r)\Big|_{r\rightarrow\infty}=-p.
\end{equation}

\subsection{Pre-buckling: relative volume change for each CS}
\label{Pre-buckling: relative volume change for each CS}

Let us consider a single CS of initial radius $S$ and   volume $V$ containing a microsphere of undeformed radius $A$ and $X=H/A$. When we increase the far-field pressure $p$ ($0<p<p_c$), this volume $V=(4/3)\pi S^3$ reduces to $v=(4/3)\pi s^3$,  where $s=S+u^m_r(S)$ denotes the deformed radius of the  CS, referring to \eqref{solution}. The relative volume change, say $\delta v$, occurring in the  pre-buckling stage is therefore given by
\begin{equation}\label{relative_change_volume_pre}
\delta v= \frac{V-v}{V}=1-\left(\frac{s}{S}\right)^3.
\end{equation}
Note that we have assumed the inner pressure inside the microsphere to remain constant under loading. This appears to be reasonable since volume changes will be small, but will not remain valid in the post-buckling regime as we shall consider in section \ref{postb}.

\subsection{Microsphere buckling}
\label{Microsphere buckling}

In this section we discuss the buckling of a spherical shell inside an unbounded elastic matrix medium, loaded by a far-field hydrostatic pressure $p$. We employ a buckling model introduced by \cite{fox-allwright01}.
Given a distribution of sizes of microspheres inside the material, our aim is to determine which of them, for a given imposed pressure $p$, have buckled and which remain unbuckled. 



In \cite{fox-allwright01} a criterion was derived for the buckling of a spherical shell embedded in an elastic material and loaded by a far-field hydrostatic pressure under the main assumptions that deformations are axisymmetric and the shell is inextensible. They also neglected the inner gas pressure, so in our model we must set $p_{\textsc{{in}}}=0$ in the pre-buckling phase. This latter simplification is, in fact, not too severe as the displacement is affected very little by internal pressure pre-buckling.
The assumption of axisymmetric buckling is not a restriction; \cite{wesolowski67} showed that the critical mode number for buckling is the same whether the eigenmode is symmetric or not. Also, for glassy shells in the present model, it is easy to show that the assumption of inextensibility is consistent with the estimates found for the radial and shear stresses.

\cite{fox-allwright01} obtained  a formula for the critical pressure  $p(\hat{X},n)$ in the form
\begin{equation}\label{F-A-criteria}
\frac{p(\hat{X},n)}{E_s}=\frac{2}{3} \frac{1+\nu_m}{1-\nu_m} \left(1+\frac{1-\nu_s}{1+\nu_m}\frac{E_m}{E_s}\frac{1}{\hat{X}}\right)\left(p_1(n) \hat{X}^3+p_2(n) \hat{X}+p_3(n)\right)
\end{equation}
 where  $\hat{X}=H/\hat{A}$ and $n$  are the shell thickness to \textit{middle} radius ratio  of the microsphere and  the mode number respectively. Note that in the Fok-Allwright approach, $n$ is a natural number greater than 1; dilatational (n=0) and rigid-body (n=1) modes are not considered. The functions $p_1, p_2$ and $p_3$ are given by
\begin{align} \label{abc}
& p_1= \left[n(n+1)-(1-\nu_s)\right]/12(1-\nu_s^2), \quad p_2= 2/\left[(n-1)(n+2)(1+\nu_s)\right]\notag\\
& p_3= E_m/E_s \frac{(2 n^3-n^2+3n+2)-\nu_m(2n^3-3n^2+5n+2)}{(n-1)^2(n+2)\left(3n+2-2\nu_m(2n+1)\right)(1+\nu_m)}.
\end{align}
The standard approach would be to specify the shell ratio $\hat{X}$ and material constants $\nu_s,\nu_m, E_s,E_m$ and substitute these into \eqref{F-A-criteria} to give the critical buckling pressure found by minimizing with respect to $n$ (and thus this also gives the corresponding buckling mode $n$). Here, however we need to approach the problem slightly differently since we have a distribution of microsphere sizes and we wish to know what the state of that distribution of microspheres (buckled/unbuckled) is at a given pressure. It will prove convenient, therefore, to treat $n$ as a parameter as we now explain.
Assuming $\hat{X}$ is for now \textit{unspecified}, by a continuity argument, assuming that $n>1$ is a given real number we determine the minimum by insisting that  $\partial p(\hat{X},n)/\partial n=0$ and solve for $\hat{X}$. From trivial algebraic considerations it is straightforward to show the existence of a minimum for $p(\hat{X},n)$ via
\begin{equation}\label{third_degree}
p_1'(n)\hat{X}^3+p_2'(n) \hat{X}+p_3'(n)=0
\end{equation}
where prime denotes differentiation with respect to argument. The real positive root $\hat{X}$  in \eqref{third_degree} depends on the mode number  $n$, which we denote by $\hat{X}_c(n)$ where the subscript $c$ refers to \textit{critical}.
Thus, we specify $n\in(1,\infty)$, determine the corresponding $\hat{X}_c$ from \eqref{third_degree}, and then the corresponding  $p_c=p(\hat{X}_c,n)$ from \eqref{F-A-criteria}. In this manner we therefore know that shells in the range $\hat{X}\leq\hat{X}_c$ are buckled whereas those for $\hat{X}>\hat{X}_c$ remain unbuckled.

\begin{figure}
\psfrag{0.5}[c][l]{\small{$0.5$\hspace{0.1cm}}}
\psfrag{1.0}[c][l]{\small{$1.0$\hspace{0.1cm}}}
\psfrag{1.5}[c][l]{\small{$1.5$\hspace{0.1cm}}}
\psfrag{2.0}[c][l]{\small{$2.0$\hspace{0.1cm}}}
\psfrag{2.5}[c][l]{\small{$2.5$\hspace{0.1cm}}}
\psfrag{0.005}[c][b]{\small{$0.005$}}
\psfrag{0.010}[c][b]{\small{$0.010$}}
\psfrag{0.015}[c][b]{\small{$0.015$}}
\psfrag{0.020}[c][b]{\small{$0.020$}}
\psfrag{0.025}[c][b]{\small{$0.025$}}
\psfrag{B}[cb][lt]{\footnotesize{$\dfrac{p_c}{\mu_m}$}}
\psfrag{A}[cl][cr]{\footnotesize{$\hat{X}_c$}}
	\centering
		\includegraphics[scale=0.9]{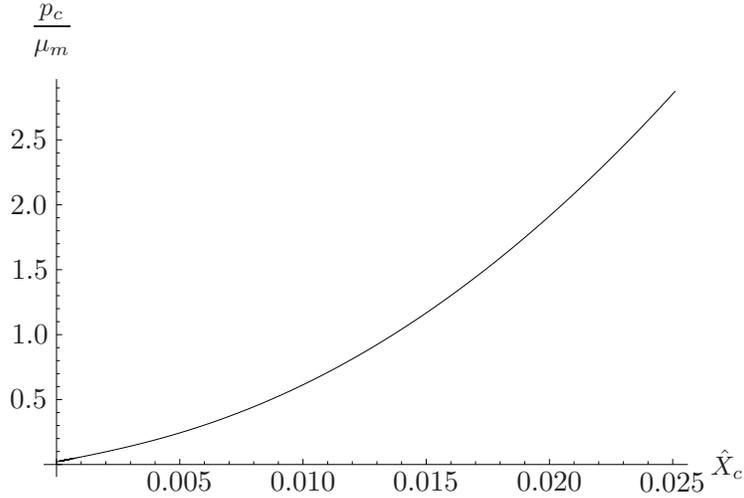}
			\caption{Plot of the critical hydrostatic pressure $p_c/\mu_m$ as a function of the critical ratio $\hat{X}_c$ predicted by the \cite{fox-allwright01} model, obtained for $n\in[13,1000]$, when we use the material constants specified in \eqref{material-constants} and we assume $p_{\textsc{{in}}}=0$.}
	\label{buckling criterions}
\end{figure}

In Fig.\ \ref{buckling criterions} we plot the predictions given by the \cite{fox-allwright01} model for the critical pressure $p_c$ as a function of critical size $\hat{X}_c$, letting the buckling mode parameter $n$ lie in the range  $[13,1000]$, assuming $p_{\textsc{{in}}}=0$ and given the material constants in \eqref{material-constants}. This range of buckling modes gives rise to realistic pressures; choosing a lower $n$ corresponds to a higher pressure.

\section{Post-buckling behaviour: nonlinear elastic response} \label{postb}
As we have emphasized before, as we increase the hydrostatic load $p$ gradually, an increasing number of shells will transition to a buckled state. When the shell buckles, there will be some complex modification to the structure of the shell and the local matrix medium. Buckling of the shell will result in a local loss of rigidity and (initially at least) a macroscopic increase in compressibility. Modelling the exact modification to the numerous shell structures is a formidable task and hence from a modelling viewpoint, in order to model the post-buckling behaviour we shall make the following simplification. For pressures $p>p_c$, for a CS region, we assume that the spherical shell region is replaced by a spherical cavity, which at the pressure $p_c$, has the same radius as the microsphere at its buckling pressure. The post-buckling stage of behaviour  will be analyzed  under the nonlinear deformation assumption, because the loss of rigidity of the microsphere will permit finite deformations if appreciable pressure $p/\mu_m$ is applied (note again Fig.\ \ref{displacement}).   We wish to understand how an additional increase in pressure in the CS region, past the buckling pressure $p_c$, decreases the volume of the CS further and subsequently we shall derive this influence on the macroscopic volume of the composite material.

From the pre-buckling analysis we can determine exactly the predictions of the deformed radii $s_c$ and $a_c$ of the CS (with corresponding initial radii $S$ and $A$ respectively) where the subscript $c$ indicates the critical value at the buckling pressure $p_c$. We can therefore begin our nonlinear analysis from those values, further increasing the load pressure until we reach the chosen load $p$. This two stage linear-nonlinear approach is however not particularly appealing; in fact we are able to consider the volume change in the post-buckling regime by considering an alternative (nonlinear) problem from the outset (i.e.\ increasing the far-field pressure from zero) that is statically equivalent to the linear problem in the pre-buckling regime as we shall now show.

\begin{figure}
\psfrag{A}{$A$}
\psfrag{ac}{$a_c$}
\psfrag{a}{$\bar{a}$}
\psfrag{p}{$p$}
\psfrag{pc}{$p_c$}
\psfrag{pin}{$p^s_{\textnormal{IN}}$}
\psfrag{I}{Increase of external pressure}
\psfrag{S}{Statically equivalent}
\psfrag{N}{Nonlinear}
\psfrag{L}{Linear}
	\centering
		\includegraphics[scale=0.65]{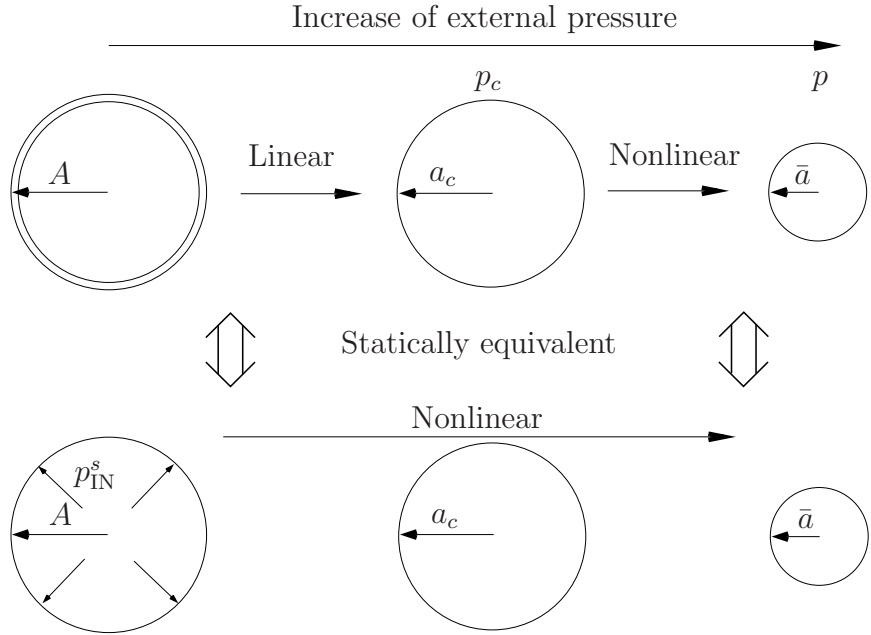}
			\caption{To determine the nonlinear deformation post-buckling we use a problem that is statically equivalent to the true problem which is drawn schematically here. Above is the linear-nonlinear process of deformation and below the fully nonlinear process. The presence of the shell is represented by the internal pressure $p_{\textnormal{IN}}^s$ in the latter, which ensures the correct radius when $p=p_c$.}
	\label{staticequiv}
\end{figure}

With reference to Fig.\ \ref{staticequiv}, consider a full nonlinear elasticity formulation of the deformation associated with an unstressed medium within which resides a \textit{cavity} with the same radius $A$ as the initially undeformed microsphere. We consider the deformation of the spherical cavity due to a far-field pressure with an additional internal `shell pressure' denoted by $p_{\textsc{{in}}}^s$ which mimics the residual presence of the shell. We choose this pressure by referring to the (linear) pre-buckling analysis, considering this to be the maximum pressure that the shell exerts on the matrix pre-buckling, i.e.\
\begin{equation}\label{newboundary-condition}
p_{\textsc{in}}^s=	-\sigma^m_{rr}(A),\ \textrm{at}\ p=p_c,
\end{equation}
such that at the critical pressure we recover to a good approximation the linear elastic solution, 
 i.e.\ to obtain  agreement for  $a_c$ and $s_c$. This therefore gives the correct starting point for volume change calculations for $p>p_c$.

We shall consider several assumptions for the nonlinear elastic matrix, and  we make use of a \textit{bar} on quantities in the post-buckling regime, in particular the undeformed radii $A,S$ and the initial volume $V$ are now denoted by $\bar{a}$, $\bar{s}$, and $\bar{v}$ respectively.

We must now determine the deformation of the cavity subject to the imposed external pressure. Since this is spherically symmetric we assume that the deformation is purely radial and we therefore write this \textit{radial deformation} in spherical polar coordinates as
\begin{equation}\label{radial-deformation_spherical}
r=r(R),\qquad \theta=\Theta,\qquad \phi=\Phi,
\end{equation}
where $(R,\Theta,\Phi)$ are the polar coordinates in the reference configuration and  $(r,\theta,\phi)$  are the polar coordinates in the current configuration, respectively, with $\textrm{d}r/\textrm{d}R>0$.
 From the result of Ericksen \citep{ericksen55} this deformation is not a controllable deformation that is possible in every \textit{compressible} homogeneous and isotropic hyperelastic material. Therefore this \textit{inhomogeneous} deformation for compressible materials has to be discussed in the context of special materials.  For example for  a special \textit{Blatz-Ko material} an analytical solution has been found in a parametric way  \citep{Chung-horgan-abeyaratne86,horgan89,horgan95zamp}. Six other classes of compressible materials have received much attention in the literature where the solution can also be found analytically \citep{carrol88,carrol91a,carrol91b,murphy92}. For an overview of such results see  \cite{horgan2001}.

In the incompressible case, thanks to the constraint of incompressibility, the radial deformation  \eqref{radial-deformation_spherical} can be treated in a much more straightforward manner and indeed is a universal solution (we refer to section 57 of \cite{truesdell-noll92} for more details). This deformation satisfies the balance equations with zero body force, its equilibrium is supported by suitable surface tractions alone, and is the same for all materials (in the class of \textit{constrained} materials). The choice of strain energy function therefore merely dictates the stress field induced by the deformation.

The special compressible solutions for radial deformation referred to above are not well suited to describe the constitutive response of a rubber-like material and therefore we choose not to use these here. Our analysis is therefore concerned firstly with purely incompressible materials before we move on to describe nearly-incompressible materials in the context of an asymptotic theory with a small parameter $\mu/\kappa\ll 1$. In the latter case we use a constitutive model proposed in the literature by \cite{horgan-murphy08} which was based on experimental behaviour (see for example \cite{penn70}).

For ease of exposition we have placed details of the theory of nonlinear elasticity associated with the subsections to follow in \ref{app:nonlin}.

%
%

\subsection{Post-buckling with incompressibility}
\label{Post-buckling with incompressibility}
The polar components of the deformation gradient associated with \eqref{radial-deformation_spherical} are given by
\begin{equation}\label{deformation_gradient}
\bm{F}= \textrm{diag} (\textrm{d}r/\textrm{d}R,r/R,r/R)
\end{equation}
and for an incompressible material  the constraint of incompressibility $\det{\bm{F}}=1$ states that
\begin{equation}\label{incompressibility_condition}
r(R)=\left(R^3+\alpha\right)^{1/3},
\end{equation}
where $\alpha=\bar{a}^3-A^3$. The equilibrium equations in the absence of body forces are $\textrm{div}{\bm{T}}=0$ where $\bm{T}$ is the Cauchy stress tensor. In the radially symmetric case these reduce to the single ordinary differential equation
\begin{equation}\label{incomp_equation}
 \dfrac{\textrm{d} T_{rr}}{\textrm{d}r}+\dfrac{2}{r}\left(T_{rr}-T_{\theta\theta}\right)=0
\end{equation}
where $\bm{T}$ is derived from a \textit{strain energy function} (SEF) $W$, as described in \ref{app:nonlin}. In this subsection we shall consider two incompressible models, the neo-Hookean material with SEF $W_{\textnormal{NH}}$ and Mooney-Rivlin material with SEF $W_{\textnormal{MR}}$, whose forms are given in \ref{app:nonlin}.

Up to the point of buckling note that we have assumed that the pressure inside the microsphere $p_{\textsc{{in}}}$ is zero since changes in volume were very small. In the post-buckling stage we require the additional pressure $p_{\textsc{{in}}}^s$ initially to maintain continuity as described in Fig.\ \ref{staticequiv} but we shall also now assume that the inner pressure $p_{\textsc{{in}}}$ can be non-constant. This is motivated by the fact that volume changes can now be large and therefore in this post-buckling stage at some point the gas interior to the cavity can be compressed so much as to act to stiffen the material. We therefore may anticipate a Boyle's law type relation for a massless ideal gas, of the form
\begin{equation}
p_{\textsc{{in}}}= p_{\textsc{{in}}}^s+p_{\textsc{{in}}}^b 
\end{equation}
where
$p_{\textsc{{in}}}^s$ is a constant that accounts for the residual effect of the buckled shell and   
\begin{equation}
p_{\textsc{{in}}}^b = p_{\textsc{{atm}}}\left(\left(\frac{A}{\bar{a}}\right)^{3\eta}-1\right). \label{boyle}
\end{equation}
In the latter expression the pressure and volume are related through a polytropic exponent relationship, for a diatomic gas, with exponent $\eta=1.4$  where $\eta$ is the heat capacity ratio. Note that all pressures are stated relative to atmospheric pressure, which motivates the form in \eqref{boyle}.

Given a SEF $W$, it is straightforward to integrate \eqref{incomp_equation} and apply the traction boundary conditions
\begin{equation}\label{new_boundary_conditions}
T_{rr}(R)\Big|_{R\rightarrow\infty}=-p,\qquad  T_{rr}(A)= -p_{\textsc{{in}}} = -p_{\textsc{{in}}}^s -p_{\textsc{{in}}}^b
\end{equation}
in order to obtain an expression linking the deformed inner radius $\bar{a}$ and the imposed pressure difference. For the \textit{neo-Hookean} case we obtain
\begin{equation}\label{neo-hookean_solution}
\frac{p-p_{\textsc{{in}}}^s -p_{\textsc{{in}}}^b}{\mu_m}=\frac{1}{2}\left(\frac{A}{\bar{a}}\right)^4+2\left(\frac{A}{\bar{a}}\right)-\frac{5}{2},
\end{equation}
whereas for the \textit{Mooney-Rivlin} model, we find
\begin{multline}\label{mooney-rivlin_solution}
\frac{p-p_{\textsc{{in}}}^s-p_{\textsc{{in}}}^b}{\mu_m}=\left(\frac{1}{2}+\gamma\right)\left[\frac{1}{2}\left(\frac{A}{\bar{a}}\right)^4+2\left(\frac{A}{\bar{a}}\right)-\frac{5}{2}\right]+\\ \left(\frac{1}{2}-\gamma\right)\left[\left(\frac{A}{\bar{a}}\right)^2-2\left(\frac{\bar{a}}{A}\right)+1\right].
\end{multline}
Note that $\gamma$ is a material constant with $-1/2 \le \gamma \le 1/2$, and when $\gamma=1/2$ in \eqref{mooney-rivlin_solution}, we recover the neo-Hookean solution  \eqref{neo-hookean_solution}.

Using the expressions \eqref{neo-hookean_solution} or  \eqref{mooney-rivlin_solution} for $\bar{a}$, from \eqref{incompressibility_condition} we can then obtain the (post-buckling) deformed radius of the CS, i.e.\ $\bar{s}=r(S)$, predicted by the neo-Hookean \eqref{neo-hookean} or Mooney-Rivlin \eqref{Mooney-Rivlin} models, respectively.

\subsection{Post-buckling with slight compressibility}
\label{Post-buckling with slight compressibility}
 A great deal of work has been presented in the literature regarding the constrained theory of elasticity (e.g.\ incompressible materials) where many solutions have been obtained in order to describe approximations to real materials. They are approximations because of course no material is in reality completely incompressible. Numerous constitutive models have been proposed in order to model the true behaviour of the material when there is a slight deviation from incompressiblity. We assume, as before, that the material is homogeneous, isotropic, and hyperelastic.  The early contribution to this theory was from \cite{Spencer70} and an application was considered by \cite{faulkner71} who considered the time dependent radial deformation of a thick walled spherical shell of almost incompressible material. More recent work has been described by Ogden \citep{ogden76,ogden78,ogden97} and Horgan and Murphy \citep{horgan-murphy07,horgan-murphy07b,horgan-murphy08}. In order to describe the linearity between pressure and volume change (assumed to hold for pressures up to $50$ MPa) summarized by  \cite{penn70} for natural rubber,  \cite{horgan-murphy08} derived several different forms of the strain energy function and we take the form $W_{\textnormal{HM}}$ in \eqref{horgan-murphy}.

In the case of slight compressibility, we have
\begin{equation}\label{compr_condition}
\epsilon = \dfrac{\mu_m}{\kappa_m}\ll 1.
\end{equation}
We then consider a regular perturbation problem, seeking asymptotic expansions in powers of $\epsilon$ for the relevant solutions, with the results obtained in section \ref{Post-buckling with incompressibility} arising as the leading order terms. Thus, we anticipate that the leading order deformation will be described by \eqref{incompressibility_condition} and we seek corrections to this, associated with the strain energy function \eqref{horgan-murphy}. We thus derive the deformed radii $\bar{a}$ and $\bar{s}$ which are slightly modified according to the slight compressibility of the matrix.


%


We assume that we can write for the deformed radial coordinate
\begin{equation}\label{sligthly_compressible_deformation}
r(R)=r_0(R)+\epsilon r_1(R)+\epsilon^2 r_2(R)+O(\epsilon^3),
\end{equation}
where
\begin{equation}
r_0(R)=\left(R^3+\alpha\right)^{1/3}
\end{equation}
is determined from the incompressible theory (equivalently $\epsilon\rightarrow0$). 

 Employing the asymptotic scheme, whose details we provide in \ref{app:nonlin} for ease of exposition, we derive the correction term $r_1(R)$ in \eqref{sligthly_compressible_deformation} as that given in \eqref{r1}. This allows us to derive the additional volume change due to the compressibility of the matrix medium.

\subsection{Post-buckling: relative volume change for each CS}
\label{Post-buckling: relative volume change for each CS}
For $p\geq p_c$, let us consider an initial composite sphere of radius $S$ containing a microsphere of initial size $X=H/A$. The relative volume change for a buckled microsphere  is now given by
\begin{equation}\label{relative_change_volume_buckled}
\delta \bar{v}= \dfrac{V-\bar{v}}{V}=1-\left(\dfrac{\bar{s}}{S}\right)^3,
\end{equation}
where $\bar{s}=r(S)$. For an incompressible matrix therefore $\bar{s}$ is evaluated via \eqref{incompressibility_condition} for Neo-Hookean or Mooney-Rivlin materials. Alternatively, for a slightly compressible matrix with strain energy function \eqref{horgan-murphy} it is evaluated via \eqref{sligthly_compressible_deformation}.

\section{Predicted pressure-relative volume change curves for the microsphere material}
\label{Relative volume change for the entire material}

\subsection{Total relative volume change for the material}

In sections \ref{Pre-buckling: relative volume change for each CS} and \ref{Post-buckling: relative volume change for each CS} we calculated the relative volume change for each composite sphere associated with the pre-buckling and post-buckling stages respectively. The choice of the former or the latter depends upon whether the value of the applied pressure is below or above the theoretical critical pressure required to buckle the microsphere of shell thickness to radius ratio $X$ in the composite sphere, according to the Fok-Allwright theory. Our principal goal is now to use these two models in order to predict the pressure-relative volume change curve for the material as a whole when there is a distribution of different shell thicknesses.

Let us introduce a probability distribution function $F(\hat{X})$ which describes the distribution of the microsphere shell thickness to radius ratios. Thus we impose the far-field hydrostatic pressure $p$ and then use the buckling model in section \ref{Microsphere buckling} to predict the critical $X_c$ below and above which buckling will and will not occur respectively. In this way we establish, at each given pressure $p$, the proportion of microspheres that are in a buckled state. Then, in order to determine the macroscopic pressure-relative volume change curve, we use \eqref{relative_change_volume_pre} and \eqref{relative_change_volume_buckled} for that proportion of microspheres that are in the pre-buckled and post-buckled states.

The relative volume change of the entire material, say  $\delta \mathcal{V}$, is therefore  given by the  sum of all of the relative volume changes in each composite sphere, the distribution of $X$ being accounted for by the probability distribution function $F(\hat{X})$. As such we write
\begin{equation}\label{rel_vol_change_entire_material}
\delta \mathcal{V}(p)=\int_0^2 \left(\delta \bar{v}(\hat{X}) \bar{\chi}(\hat{X})+ \delta v(\hat{X}) \chi(\hat{X})\right) F(\hat{X})\ \textrm{d}\hat{X}.
\end{equation}
where $\chi(\hat{X})=1-\bar{\chi}(\hat{X})$ is the indicator function, defined as
\begin{align}
\chi(\hat{X})=\begin{cases}
0, & \hat{X}\in[0,\hat{X}_c], \\
1, & \hat{X}\in[\hat{X}_c,2].
\end{cases}
\end{align}
Note that in \eqref{rel_vol_change_entire_material} we have allowed $\hat{X}$ to take on all possible values $\hat{X}\in[0,2]$ corresponding to $H\in[0,A]$. However, we note that the buckling theory above is applicable to thin shells only and therefore the choice of $F$ is important in order for \eqref{rel_vol_change_entire_material} to give an accurate prediction. In reality the microsphere shells are very thin (O(0.01)) and thus the distribution function $F$ must model this, as we now describe.

\subsection{Parameter studies}
\label{Some predicted pressure-volume curves}

Let us now consider how the \textit{pressure-relative volume change} curve is affected by the numerous parameters in the problem and in particular we wish to understand the \textit{sensitivity} of the curve to these parameters. We first consider the form of the probability distribution function $F(\hat{X})$. From \cite{PBStech} it appears that the microsphere shell to radius ratio distribution can be described well by a Gamma distribution. We thus define
\begin{equation}\label{gamma}
F(\hat{X})= \left(\dfrac{k}{\hat{X}_0}\right)^k     \dfrac{\hat{X}^{k-1}}{\Gamma(k)}  \exp\left[-\left(k/\hat{X}_0\right)\hat{X}\right],
\end{equation}
 where $k>0$ is the shape parameter, $\hat{X}_0>0$ is the  mean value (expectation) of $\hat{X}$ and  $\Gamma(k)$ is  the Gamma function evaluated at $k$.

Parameter studies will involve choosing values for the elastic properties ($\kappa_s,\kappa_m,\mu_s,\mu_m$), the initial volume fraction $\Phi$ of microspheres, the parameters $k$ and $\hat{X}_0$ in \eqref{gamma} and the constitutive model for the nonlinear elastic matrix described in section \ref{postb}. Additionally we must decide whether or not to incorporate Boyle's law inside the microspheres during compression. This parameter set can therefore be chosen in many ways. We select certain cases in order to illustrate specific aspects of the model and hence test the sensitivity of the results to the particular parameters.

%
%


\subsubsection{Influence of nonlinear elastic constitutive model} \label{infnonlin}

\begin{figure}
\centering
\psfrag{0.2}[c][b]{\small{$0.2$}}
\psfrag{0.4}[c][b]{\small{$0.4$}}
\psfrag{0.6}[c][b]{\small{$0.6$}}
\psfrag{0.8}[c][b]{\small{$0.8$}}
\psfrag{0.002}[c][l]{\small{$0.002\hspace{0.1cm}$}}
\psfrag{0.004}[c][l]{\small{$0.004\hspace{0.1cm}$}}
\psfrag{0.006}[c][l]{\small{$0.006\hspace{0.1cm}$}}
\psfrag{0.008}[c][l]{\small{$0.008\hspace{0.1cm}$}}
\psfrag{0.010}[c][l]{\small{$0.010\hspace{0.1cm}$}}
\psfrag{0.012}[c][l]{\small{$0.012\hspace{0.1cm}$}}
\psfrag{0.014}[c][l]{\small{$0.014\hspace{0.1cm}$}}
\psfrag{0.0005}[c][c]{\small{$0.5$}}
\psfrag{0.0010}[c][c]{\small{$1.0$}}
\psfrag{0.0015}[c][c]{\small{$1.5$}}
\psfrag{0.0020}[c][c]{\small{$2.0$}}
\psfrag{0.0025}[c][c]{\small{$2.5$}}
\psfrag{0.0030}[c][c]{\small{$3.0$}}
\psfrag{B}{\footnotesize{$\delta \mathcal{V}$}}
\psfrag{A}{\footnotesize{$\dfrac{p}{\mu_m}$}}
\psfrag{D}{\footnotesize{$\mathcal{D}\times 10^3$}}
\psfrag{C}{\footnotesize{$\dfrac{p}{\mu_m}$}}
\includegraphics[scale=0.60]{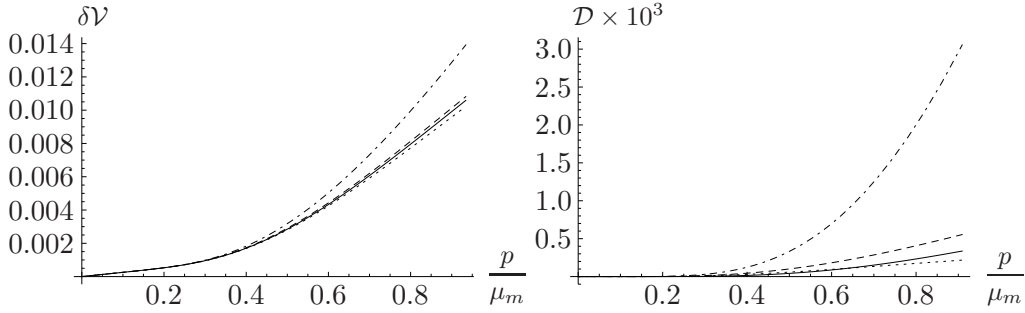}
\caption{Left: Predicted pressure-relative volume change curve with fixed parameters $\Phi=0.05$, $\mu_s,\kappa_s,\mu_m,\kappa_m$ as given in \eqref{material-constants} and $F(\hat{X})$  as in \eqref{gamma} with $\hat{X}_0=0.01,k=8$. Nonlinear model for the matrix is taken as neo-Hookean (dotted), Mooney-Rivlin  (solid) and slightly-compressible Horgan Murphy (dashed) with $\gamma=1/18$ and for reference we also plot the response when the post-buckling regime is determined via \textit{linear} elasticity (dot-dash). Right: Plot of the difference $\mathcal{D}$ in the prediction of $\delta \mathcal{V}$ via the alternative nonlinear models. $\mathcal{D}=\delta \mathcal{V}_{MR}-\delta \mathcal{V}_{NH}$ (solid),  $\mathcal{D}=\delta \mathcal{V}_{HM}-\delta \mathcal{V}_{MR}$ (dotted), $\mathcal{D}=\delta \mathcal{V}_{HM}-\delta \mathcal{V}_{NH}$ (dashed) and $\mathcal{D}=\delta \mathcal{V}_{LIN}-\delta \mathcal{V}_{MR}$ (dot-dashed).}
\label{firstplot}
\end{figure}

We start by fixing the material properties as those given in \eqref{material-constants} and we take an initial volume fraction of microspheres as $\Phi=0.05$.  Furthermore, within the distribution function $F(\hat{X})$ given in \eqref{gamma} we take the shape parameter $k=8$ as given in \cite{PBStech} and we consider a mean value $\hat{X}_0=0.01$.   Finally, we assume that the gas inside the microspheres remains at a constant atmospheric pressure, i.e.\ $p_{\textsc{{in}}}^b=0$. On the left of Fig.\ \ref{firstplot} we plot the predicted pressure-relative volume change curve for three different nonlinear elastic constitutive models in the post-buckling regime: Neo-Hookean (dotted), Mooney-Rivlin with $\gamma=1/18$ (solid) and slightly compressible Horgan-Murphy (dashed) as well as a purely linear elastic deformation (dot-dashed).

As can be seen from the left of Fig.\ \ref{firstplot}, all curves are strongly nonlinear, exhibiting a softening behaviour under loading after the initial linear elastic behaviour pre-buckling at small pressures. As should be expected all four curves are initially identical with a slight modification in the post-buckling regime where the linear or nonlinear elastic model becomes important. However even in this regime the curve is fairly insensitive to the constitutive model employed, nonlinear models yielding almost identical results. The linear model is fairly close to the nonlinear predictions at these pressures but as pressure increases the linear results depart significantly as should be expected from this approximate theory.
%

In order to better compare the models, on the right of Fig.\ \ref{firstplot} we plot the \textit{difference} between the relative volume change predicted by the different elastic post-buckling models. For example the difference between Horgan-Murphy and Mooney-Rivlin is calculated as $\mathcal{D}=\delta \mathcal{V}_{HM}-\delta \mathcal{V}_{MR}$ and analogously for the other two possiblities. Given the scale, it is clear that the difference between any of the nonlinear models is $O(10^{-4})$. We reiterate that at these pressures the linear elastic model is relatively close to the nonlinear models but at higher pressures there is a significant departure.

\begin{figure}
\centering
\psfrag{5}[c][b]{\small{$5$}}
\psfrag{10}[c][b]{\small{$10$}}
\psfrag{15}[c][b]{\small{$15$}}
\psfrag{20}[c][b]{\small{$20$}}
\psfrag{25}[c][b]{\small{$25$}}
\psfrag{0.01}[c][l]{\small{$0.01$}}
\psfrag{0.02}[c][l]{\small{$0.02$}}
\psfrag{0.03}[c][l]{\small{$0.03$}}
\psfrag{0.04}[c][l]{\small{$0.04$}}
\psfrag{0.05}[c][l]{\small{$0.05$}}
\psfrag{B}{\footnotesize{$\delta \mathcal{V}$}}
\psfrag{A}{\footnotesize{$\dfrac{p}{\mu_m}$}}
\includegraphics[scale=0.75]{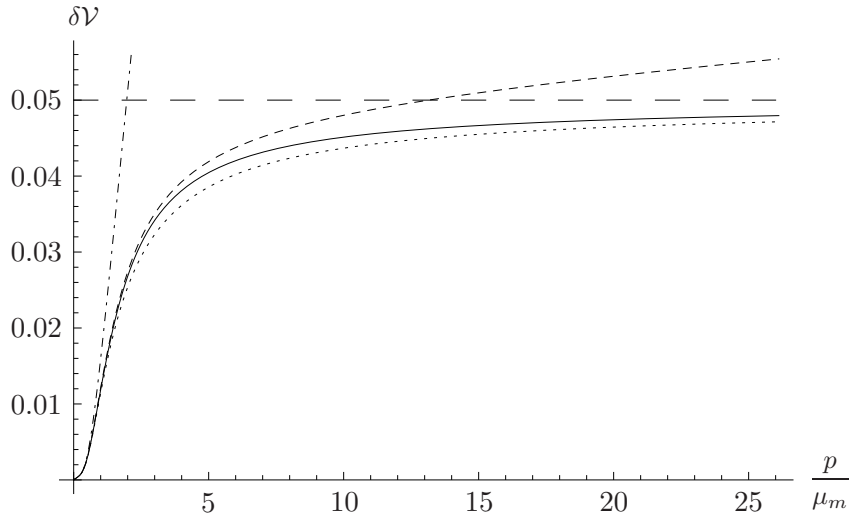}
\caption{Prediction of $\delta \mathcal{V}$ for large values of $p/\mu_m$ with all parameters as given in Fig. \ref{firstplot} and plotted for Neo-Hookean  (dotted), Mooney-Rivlin  (solid) and slightly-compressible Horgan Murphy (dashed) nonlinear models as well as a linear elastic prediction (dot-dashed).}
\label{nonlinearmodelbis}
\end{figure}

Although here we are predominantly interested in $p/\mu_m=O(1)$ we note with reference to Fig.\ \ref{nonlinearmodelbis}, that for both \textit{incompressible} nonlinear materials, as
$p/\mu_m\rightarrow\infty$, $\delta \mathcal{V}\rightarrow 0.05$, as we should expect in that case (parameters chosen are those associated with Fig.\ \ref{firstplot}). The Horgan-Murphy model (dashed) predicts a slightly different limit for $\delta \mathcal{V}$ as shown in Fig.\ \ref{nonlinearmodelbis} since this takes into account the slight compressibility of the matrix. The curve associated with the linear elastic model illustrates its unrealistic nature for large deformations.

\subsubsection{Influence of volume fraction, shell properties and pressure law}

\begin{figure}
\centering
\psfrag{0.2}[c][b]{\small{$0.2$}}
\psfrag{0.4}[c][b]{\small{$0.4$}}
\psfrag{0.6}[c][b]{\small{$0.6$}}
\psfrag{0.8}[c][b]{\small{$0.8$}}
\psfrag{0.000}[c][l]{\small{$$}}
\psfrag{0.005}[c][l]{\small{$0.005$}}
\psfrag{0.010}[c][l]{\small{$0.010$}}
\psfrag{0.015}[c][l]{\small{$0.015$}}
\psfrag{0.020}[c][l]{\small{$0.020$}}
\psfrag{A}{\footnotesize{$\dfrac{p}{\mu_m}$}}
\psfrag{B}{\footnotesize{$\delta \mathcal{V}$}}
\includegraphics[scale=0.65]{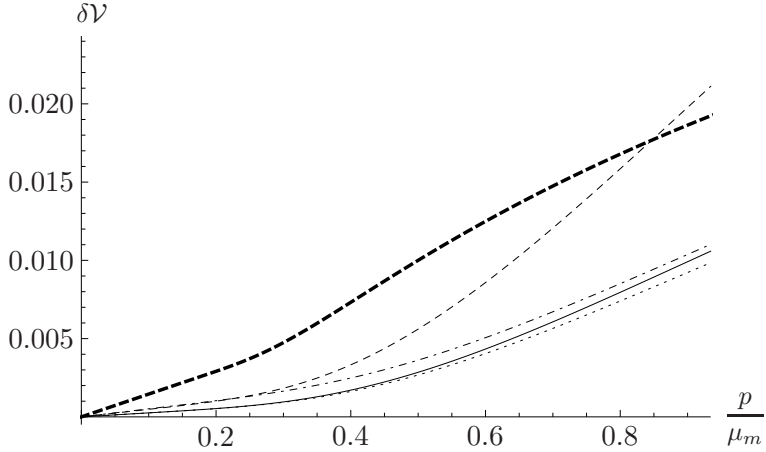}
\caption{Parameter study for the predicted pressure-relative volume change curve. The solid curve corresponds to the parameters assumed in Fig.\ \ref{firstplot} with a Mooney-Rivlin matrix medium. We then vary one specific aspect of the model to assess sensitivity. Different curves correspond to: increasing microsphere volume fraction from $\Phi=0.05$ to $\Phi=0.1$ (dashed); incorporating Boyle's law \eqref{boyle} for the gas interior to the microsphere instead of constant pressure (dotted);
softer shell properties given by \eqref{shell_softer} (thick dashed); slightly more compressible matrix phase given by
\eqref{host_softer} (dot-dashed).}
\label{secondplot}
\end{figure}

We concluded in section \ref{infnonlin} that the predicted curves are relatively insensitive to the nonlinear elastic model employed. As a result of this insensitivity, let us now model the matrix as an incompressible Mooney-Rivlin medium - a standard model for a rubber-like matrix medium. Thus, with the solid curve from Fig. \ref{firstplot} as a starting point, also plotted is a reference curve in Fig.\ \ref{secondplot}, and let us vary other parameters in order to assess their influence. For each curve we keep all parameters fixed except one control parameter in order to assess its particular effect. Thus, the dashed curve in Fig.\ \ref{secondplot} corresponds to changing only the volume fraction of microspheres from $\Phi=0.05$ to $\Phi=0.1$. The dotted curve corresponds to incorporating Boyle's law \eqref{boyle} for the gas interior to the microsphere instead of constant pressure. The thick dashed curve corresponds to softer shell material properties, i.e.\
\begin{equation}\label{shell_softer}
\mu_s= 0.126\ \textrm{GPa},\quad \kappa_s = 0.21\ \textrm{GPa},
\end{equation}
and the dot-dashed curve is associated with slightly more compressible matrix material properties (for the linear elastic pre-buckling portion of the curve), i.e.\
\begin{equation}\label{host_softer}
\mu_m =1.2\  \textrm{MPa},\quad \kappa_m = 0.4\ \textrm{GPa}.
\end{equation}
These correspond to a Poisson's ratio of $\nu_m=0.4985$.

Let us assess each one of these in turn. Increasing the microsphere volume fraction, whilst keeping their distribution fixed has the expected effect: the curve remains qualitatively similar, becoming relatively softer in the nonlinear regime. Incorporating Boyle's law should add some stiffness in the post-buckling regime and this can be seen in the figure; but its effect is rather modest. The dotted and solid curves are identical until the post-buckling effects become important, around $p/\mu_m=0.2$, and then increased pressure interior to the microsphere does yield a small additional stiffness. Softer shell properties are expected to have a larger effect in the transition region from pre to post buckling as the shells will clearly buckle at lower pressures. This can clearly be seen in the figure; an order of magnitude change to the properties has modified the curve significantly. However, the insensitivity post-buckling can be seen by virtue of this curve and the solid curve remaining parallel in this regime. Finally, the small decrease in matrix Poisson's ratio corresponding to the dot-dashed curve yields the expected effect; the material becomes slightly softer in the linear region, recovering an identical nonlinear response to the reference Mooney-Rivlin case in the post-buckling region.

As perhaps should be expected, there is a great sensitivity to the material properties of the shell, but the model is relatively insensitive to other parameters. 

\subsubsection{Influence of probability distribution function parameters}

\begin{figure}
\centering
\psfrag{0.2}[c][b]{\footnotesize{$0.2$}}
\psfrag{0.4}[c][b]{\footnotesize{$0.4$}}
\psfrag{0.6}[c][b]{\footnotesize{$0.6$}}
\psfrag{0.8}[c][b]{\footnotesize{$0.8$}}
\psfrag{0.01}[c][b]{\footnotesize{$0.01$}}
\psfrag{0.02}[c][b]{\footnotesize{$$}}
\psfrag{0.03}[c][b]{\footnotesize{$0.03$}}
\psfrag{0.04}[c][b]{\footnotesize{$$}}
\psfrag{0.05}[c][b]{\footnotesize{$0.05$}}
\psfrag{0.06}[c][b]{\footnotesize{$$}}
\psfrag{50}[c][l]{\footnotesize{$50$}\hspace{-0.1cm}}
\psfrag{100}[c][l]{\footnotesize{$100$}\hspace{-0.1cm}}
\psfrag{150}[c][l]{\footnotesize{$150$}\hspace{-0.1cm}}
\psfrag{200}[c][l]{\footnotesize{$200$}\hspace{-0.1cm}}
\psfrag{0}[c][l]{\small{$$}}
\psfrag{2}[c][l]{\footnotesize{$0.002$}\hspace{0.6cm}}
\psfrag{4}[c][l]{\footnotesize{$0.004$}\hspace{0.6cm}}
\psfrag{6}[c][l]{\footnotesize{$0.006$}\hspace{0.6cm}}
\psfrag{8}[c][l]{\footnotesize{$0.008$}\hspace{0.6cm}}
\psfrag{10}[c][l]{\footnotesize{$0.010$}\hspace{0.4cm}}
\psfrag{5}[c][l]{\footnotesize{$0.005$}\hspace{0.6cm}}
\psfrag{15}[c][l]{\footnotesize{$0.015$}\hspace{0.4cm}}
\psfrag{20}[c][l]{\footnotesize{$0.020$}\hspace{0.4cm}}
\psfrag{0.005}[c][b]{\footnotesize{$0.005$}}
\psfrag{0.010}[c][b]{\footnotesize{$$}}
\psfrag{0.015}[c][b]{\footnotesize{$0.015$}}
\psfrag{0.020}[c][b]{\footnotesize{$$}}
\psfrag{0.025}[c][b]{\footnotesize{$0.025$}}
\psfrag{0.030}[c][b]{\footnotesize{$$}}
\psfrag{B}{\footnotesize{$F$}}
\psfrag{A}{\footnotesize{$\hat{X}$}}
\psfrag{D}{\footnotesize{$\delta \mathcal{V}$}}
\psfrag{C}{\footnotesize{$\dfrac{p}{\mu_m}$}}
\includegraphics[scale=0.65]{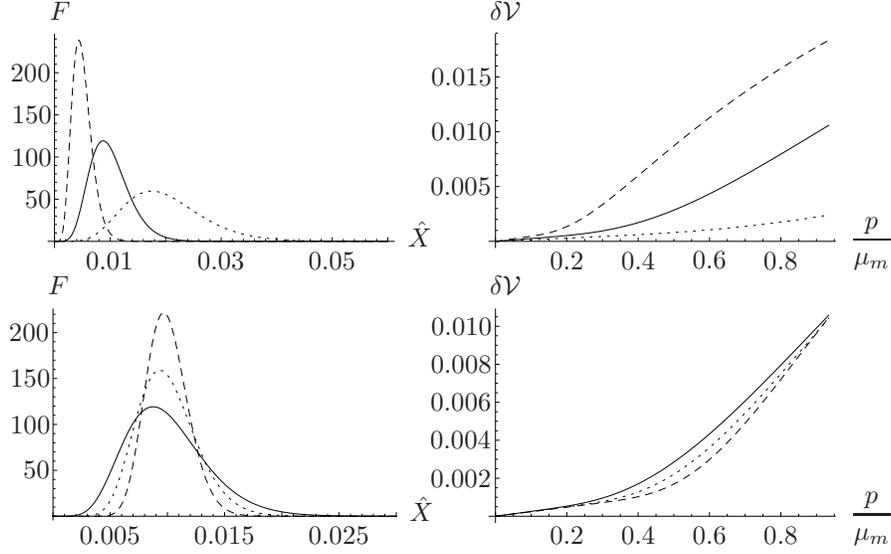}
\caption{Plots of the distribution function (left) against shell thickness to radius ratio and corresponding pressure-relative volume change curves (right). The parameter $k=8$ is held fixed and we vary $\hat{X}_0$ (top) whereas we keep $\hat{X}_0=0.01$ fixed and vary $k$ (bottom). Specifically we take $\hat{X}_0=0.01, 0.02$ and $0.005$ (solid, dotted and dashed respectively) (top) and $k=8,15$ and $30$ (solid, dotted and dashed respectively) (bottom).}
\label{thirdplot}
\end{figure}

Let us now take the Mooney-Rivlin reference curve (solid) as plotted in Fig. \ref{firstplot} and vary the distribution function parameters 
in $F$ from those of the reference material $\hat{X}_0=0.01$ and $k=8$. In Fig.\ \ref{thirdplot} we plot the distribution function (left) and corresponding pressure-relative volume change curves (right) whilst keeping $k=8$ fixed and varying $\hat{X}_0$ (top) and then keeping $\hat{X}_0=0.01$ fixed and varying $k$ (bottom). As perhaps should be expected there is great sensitivity to the average shell thickness to radius ratio. This can be seen by the large modifications to the pressure-volume curves in the top-right figure. The main effect of varying $k$ is the later (in terms of higher pressure) softening of the material, although its influence is less marked than variation in $\hat{X}_0$. This information is useful from the viewpoint of knowing the correct type and distribution of microspheres to use in the composite.

%
%

%
%
%
%

\begin{figure}
\centering
\psfrag{50}[c][l]{\footnotesize{$50$}}
\psfrag{100}[c][l]{\footnotesize{$100$}}
\psfrag{150}[c][l]{\footnotesize{$150$}}
\psfrag{200}[c][l]{\footnotesize{$200$}}
\psfrag{250}[c][l]{\footnotesize{$250$}}
\psfrag{300}[c][l]{\footnotesize{$300$}}
\psfrag{0}[c][l]{\small{$$}}
\psfrag{0.005}[c][b]{\footnotesize{$0.005$}}
\psfrag{0.010}[c][b]{\footnotesize{$$}}
\psfrag{0.015}[c][b]{\footnotesize{$0.015$}}
\psfrag{0.020}[c][b]{\footnotesize{$$}}
\psfrag{0.025}[c][b]{\footnotesize{$0.025$}}
\psfrag{0.030}[c][b]{\footnotesize{$$}}
\psfrag{0.2}[c][b]{\footnotesize{$0.2$}}
\psfrag{0.4}[c][b]{\footnotesize{$0.4$}}
\psfrag{0.6}[c][b]{\footnotesize{$0.6$}}
\psfrag{0.8}[c][b]{\footnotesize{$0.8$}}
\psfrag{2}[c][l]{\footnotesize{$0.002$}\hspace{0.65cm}}
\psfrag{4}[c][l]{\footnotesize{$0.004$}\hspace{0.65cm}}
\psfrag{6}[c][l]{\footnotesize{$0.006$}\hspace{0.65cm}}
\psfrag{8}[c][l]{\footnotesize{$0.008$}\hspace{0.65cm}}
\psfrag{10}[c][l]{\footnotesize{$0.010$}\hspace{0.5cm}}
\psfrag{B}{\footnotesize{$F$}}
\psfrag{A}{\footnotesize{$\hat{X}$}}
\psfrag{D}{\footnotesize{$\delta \mathcal{V}$}}
\psfrag{C}{\footnotesize{$\dfrac{p}{\mu_m}$}}
\includegraphics[scale=0.6]{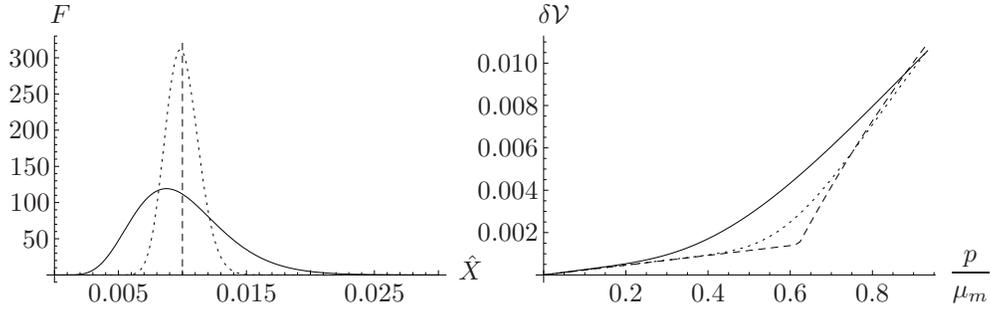}
\caption{Illustrating how the shape of the load curve is modified in the limit $k\rightarrow\infty$. We choose $\hat{X}_0=0.01$ and take $k=8$ (solid), $k=50$ (dotted) and the limit as $k\rightarrow\infty$ (dashed). The limiting case when there is only one size of microsphere shell in the composite ($k\rightarrow\infty$) is reflected in the shape of the pressure-volume curve illustrated on the right, i.e.\ the appearance of a discontinuity in derivative of the curve. The curve becomes smoother in this region as $k$ is progressively reduced.}
\label{diraclimit}
\end{figure}

Finally, we illustrate how the `kink' in the load curve is associated primarily with the distribution function of the shell thicknesses. Let us choose $\hat{X}_0=0.01$ and successively increase $k$ which has the effect of tending the distribution function towards a dirac delta function as can be seen in Fig.\ \ref{diraclimit} where we take $k=8$ (solid), $k=50$ (dotted) and the limit as $k\rightarrow\infty$ (dashed). The limiting case when there is only one size of microsphere shell in the composite ($k\rightarrow\infty$) is reflected in the shape of the pressure-volume curve illustrated on the right of Fig.\ \ref{diraclimit}, manifested by a discontinuity in derivative of the curve. When $k$ is made finite and reduced the curve becomes progressively smoother.

\section{Conclusions}

We have presented a model that predicts the nonlinear \textit{pressure-relative volume change} loading curve associated with a microsphere elastomeric composite material. The nonlinearity is induced by several mechanisms: (i) incorporating a distribution of sizes of spherical shells which thus buckle successively according to the applied load, (ii) modelling the post-buckling behaviour of the matrix as a nonlinear elastic material, (iii) incorporating Boyle's law for the pressure interior to the microsphere in the post-buckling regime.

In this initial study we have neglected any interaction between microspheres both in terms of the buckling analysis and the determination of the change in volume of the composite. Therefore we anticipate that the model is valid only for low volume fractions of microspheres.

Parameter studies reveal that, although it appears important to include nonlinear behaviour in the post-buckling stage, the curves are largely insensitive to the chosen nonlinear elastic model. However the curves \textit{are} particularly sensitive to the properties of the microsphere, including shell properties and the distribution of shell thickness, particularly the choice of mean shell thickness to radius ratio $\hat{X}_0$. Furthermore, the shape of the `kink' in the load curve is associated primarily with the distribution function of the shell thicknesses. The smaller $k$ is, the smoother the transition to the nonlinear post-buckling state.

We have also seen that when Boyle's law is incorporated in the post-buckling regime (the influence of a gas inside the shell pre-buckling is
negligible as so is omitted in the model) there is competition between the softening of the material due to microsphere buckling and stiffening due to Boyle's law. It transpired that the latter contribution is small, as is visible in the transition from pre to post buckling in Fig.\ \ref{secondplot}. The near-incompressible nature of the matrix medium dictates that eventually the response of the composite will change, as the pressure increases, from an initial softening material to one which stiffens for large pressures ($p/\mu_m>O(1)$) as is seen in Fig.\ \ref{nonlinearmodelbis}.

Note that in order to solve the post-buckling problem of a single spherical cavity embedded in a unbounded medium subject to inner and external hydrostatic pressure we used the theory of almost-incompressible materials, posing an asymptotic expansion for the Horgan-Murphy model \eqref{horgan-murphy} (see \cite{horgan-murphy08}).

Follow-on work will consider the accuracy of the buckling model, by comparing with alternative (e.g.\ \cite{Jones-Chapman-Allwright07}) and new models. We shall also consider the effect of the interaction of microspheres on buckling. This is clearly important when volume fractions of microspheres become larger, a common case in practice.

\subsection*{Acknowledgements}

The authors are grateful to the Engineering and Physical Sciences Research Council for funding this work via grant EP/H050779/1. They are also grateful to Dr Philip Cotterill and Dr Peter Brazier-Smith (Thales Underwater Systems Ltd) and Dr John Smith (DSTL) for their assistance regarding various aspects of this work. The authors are also grateful to Professor Bing Li (Technical Institute of Physics and Chemistry, Chinese Academy of Science) and Dr James Busfield (Queen Mary, University of London) for their willingness to provide figures \ref{fig:HGM} and \ref{shorterimage} respectively, in order to reproduce them here.

\appendix
\section{Nonlinear elasticity theory} \label{app:nonlin}

\subsection{Incompressible theory}

Given the deformation gradient $\bm{F}$, we compute the physical components of the left Cauchy-Green strain tensor $\bm{B}\equiv\bm{F}\bm{F}^T$ from \eqref{deformation_gradient} and find its first three principal invariants from
\begin{equation}\label{invariants_incompr}
I_1=\mathrm{tr} \bm{B},\quad I_2=\frac{1}{2}[(\mathrm{tr} \bm{B})^2-\mathrm{tr} \bm{B}^2],\quad I_3=\mathrm{det} \bm{B}.
\end{equation}
For an isotropic incompressible hyperelastic solid, the Cauchy stress tensor $\bm{T}$ is then related to the strain via
\begin{equation} \label{T_after_CH}
\bm{T}=-q\bm{I}+2 W_1\bm{B}-2W_2\bm{B}^{-1},
\end{equation}
where $q$ is the Lagrange  multiplier introduced by the incompressibility constraint, $W(I_1,I_2)$ is the strain-energy density and
$W_i\equiv\partial W/ \partial I_i$.

In the incompressible case, in order to model the nonlinear constitutive response of the matrix material, let us consider two alternative strain energy functions, the so-called neo-Hookean and Mooney-Rivlin models:
\begin{align}
W_{\textnormal{NH}}&=\dfrac{\mu_m}{2} (I_1-3), \label{neo-hookean}\\
W_{\textnormal{MR}}&=\frac{1}{2}\left(\frac{1}{2}+\gamma\right)\mu_m(I_1-3)+\frac{1}{2}\left(\frac{1}{2}-\gamma\right)\mu_m(I_2-3),\label{Mooney-Rivlin}
\end{align}
where $\mu_m>0$ is the shear modulus of the matrix (as introduced above) for infinitesimal deformations and $\gamma$ is a non-dimensional constant in the range $-1/2\leq\gamma\leq1/2$.

\subsection{Slightly-compressible theory}

In the case of generally compressible materials, the Cauchy stress can be written as
\begin{equation} \label{T_Cayley-Hamilton}
\bm{T}=\beta_0 \bm{I}+\beta_1\bm{B}+\beta_{-1} \bm{B}^{-1},
\end{equation}
where
\begin{align}
&\beta_0(I_1,I_2, I_3)=\frac{2}{\sqrt{I_3}} \left[I_2\frac{\partial W}{\partial I_2}+I_3 \frac{\partial W}{\partial I_3}\right], \nonumber \\
& \beta_{1}(I_1,I_2, I_3)=\frac{2}{\sqrt{I_3}} \frac{\partial W}{\partial I_1}, \label{elastic_response_functions}  \\
&\beta_{-1}(I_1,I_2, I_3)=-2 \sqrt{I_3}\frac{\partial W}{\partial I_2}. \nonumber
\end{align}
A variety of strain energy functions can be proposed, but one in particular was described by \cite{horgan-murphy08} in the form
\begin{equation}\label{horgan-murphy}
W_{\textnormal{HM}}= \frac{\mu_m}{2}\left(\frac{1}{2}+\gamma\right)(I_1-3 I_3^{1/3})+\frac{\mu_m}{2}\left(\frac{1}{2}-\gamma\right)(I_2-3 I_3^{2/3})+\frac{\kappa_m}{2} (I_3^{1/2}-1)^2
\end{equation}
where $\gamma$ is an arbitrary constant, $\mu_m$ is the infinitesimal shear modulus and $\kappa_m$ is the infinitesimal bulk modulus. The latter two material parameters are usually related for an almost incompressible (slightly compressible) material through the additional assumption
\begin{equation}\label{compr_condition2}
\epsilon= \dfrac{\mu_m}{\kappa_m}\ll 1.
\end{equation}
Note from  \eqref{horgan-murphy} that $W_{\textnormal{HM}}$ can be considered as a slightly compressible generalization of the incompressible Mooney-Rivlin strain energy function \eqref{Mooney-Rivlin} since
\begin{equation}
W_{\textnormal{HM}}(I_1,I_2,I_3=1)=W_{\textnormal{MR}}(I_1,I_2).
\end{equation}

We proceed to solve the boundary value problem by posing the radial displacement as a regular asymptotic expansion in powers of $\epsilon$, i.e.\
\begin{equation}\label{sligthly_compressible_deformation2}
r(R)=r_0(R)+\epsilon r_1(R)+\epsilon^2 r_2(R)+O(\epsilon^3),
\end{equation}
where
\begin{equation}
r_0(R)=\left(R^3+\alpha\right)^{1/3}
\end{equation}
is determined from the incompressible theory (equivalently $\epsilon\rightarrow0$). The correction terms $r_1(R)$ and $r_2(R)$ are the same order of magnitude as $r_0(R)-R$. Making use of the unconstrained theory our goal is to determine $r_1(R)$ and in order to do this it is necessary to retain terms of $O(\epsilon^2)$ in \eqref{sligthly_compressible_deformation2}.

Let us expand the Cauchy stress tensor $\bm{T}$, whose form for the compressible problem is given in \eqref{T_Cayley-Hamilton}, in the form
\begin{equation}\label{expansion_stress}
\bm{T}=\bm{T}_0+\epsilon \bm{T}_1+O(\epsilon^2).
\end{equation}
Since $\beta_0$ in  \eqref{elastic_response_functions} involves the term $I_3 \partial W/\partial I_3$, the hydrostatic part of  $\bm{T}_0$ involves the term $r_1$ (see \cite{Spencer70}) which is then determined in a manner described below.  We write all other equations in the form of asymptotic expansions. In particular we note that
\begin{equation}
\textrm{div}{\bm{T}}=\textrm{div}{\bm{T}}_0+\epsilon \hspace{0.1cm}\textrm{div}{\bm{T}}_1+0(\epsilon^2),
\end{equation}
and the traction boundary conditions must be written as
\begin{equation}\label{new_boundary_conditions_nearlyincomp}
T_{rr}(R)_{|R\rightarrow\infty}=-p+0\epsilon+O(\epsilon^2),\qquad  T_{rr}(A)=-p_{\textsc{{in}}}^s -p_{\textsc{{in}}}^b+0\epsilon+O(\epsilon^2),
\end{equation}
noting that they give inhomogeneous conditions at leading order and homogeneous conditions at higher orders.
Using these we can then equate terms at each order in $\epsilon$ and solve the resulting problems.

Using the boundary condition \eqref{new_boundary_conditions_nearlyincomp}$_1$  to leading order and the equilibrium  equation $\textrm{div}{\bm{T}}_0=0$  we obtain  the general expression for $T_{0rr}$,
\begin{equation}\label{radial_stress_leading}
T_{0rr}(R)=-p-2\int^{\infty}_R  \dfrac{\textrm{d} r_0(\rho)}{\textrm{d} \rho } \dfrac{T_{0rr}(\rho)-T_{0\theta\theta}(\rho)}{r_0(\rho)}\ \textrm{d}\rho.
\end{equation}
Imposing the leading order boundary condition \eqref{new_boundary_conditions_nearlyincomp}$_2$ then allows us to recover the relationship \eqref{mooney-rivlin_solution}, i.e.\ the result associated with the incompressible solution \eqref{incompressibility_condition}.

Next equating \eqref{radial_stress_leading} with the zeroth order term $T_{0rr}$ found in \eqref{expansion_stress},  allows us to find a  linear first order ordinary differential equation for $r_1$, whose solution is given explicitly by
\begin{equation}
r_1(R)=\frac{2 R^2 r_0 ^2(-1+2 \gamma )+R^4 (1+2 \gamma )+R^3r_0 \left(1-6 \gamma -\frac{4 p}{3 \mu_m }\right)+4 C_1r_0}{4 r_0^{^3}}, \label{r1}
\end{equation}
where $C_1$ is an integration constant to be determined. In order to determine $C_1$  we must continue our analysis to the next order i.e.\ we solve $\textrm{div}{\bm{T}}_1=0$.  The first order balance equation is a linear second order differential equation for the unknown function $r_2(R)$ whose solution we do not reproduce here (for brevity, but also because we are primarily interested in $r_1(R)$). The function $r_2(R)$ involves two constants of integration which we shall call $C_2$ and $C_3$. It is straightforward to determine that only one of these constants, say $C_2$, is involved in the component  $T_{1rr}$ of the stress, in addition to $C_1$. Using the $O(\epsilon)$ boundary conditions \eqref{new_boundary_conditions_nearlyincomp} for the stress $\bm{T}_1$, it is then possible to determine both $C_1$ and $C_2$. In particular since we are interested in only the first order correction to $r_0$ we state only $C_1$ which takes the form
\begin{equation}
C_1= -\dfrac{\alpha_8 \bar{a}^8+\alpha_7 \bar{a}^7+\alpha_6 \bar{a}^6+\alpha_5 \bar{a}^5+\alpha_4 \bar{a}^4+\alpha_3 \bar{a}^3+\alpha_2 \bar{a}^2+\alpha_1 \bar{a}+\alpha_0}{\Big[480 \bar{a} \left(A^3+\bar{a}^3\right) \left(\bar{a}^2 (1-2 \gamma )+A^2 (1+2 \gamma )\right) \mu_m \Big]}
\end{equation}
where
\begin{align}
\alpha_8 &=A (40 p (7+6 \gamma )-3 (83+4 \gamma  (43 \gamma-57)) \mu_m ),  \notag \\
\alpha_7 &=240 A^2 (1-2 \gamma )^2 \mu_m, \notag \\
\alpha_6 &=-20 A^3 (2 \gamma-1) (4 p+3 (6 \gamma-1) \mu_m ),  \notag \\
\alpha_5 &=120 A^4 \left(4 \gamma ^2-1\right) \mu_m, \notag \\
\alpha_4 &=-10 A^5 (p (4+8 \gamma )+3 (5+4 \gamma  (9 \gamma-5)) \mu_m ), \notag \\
\alpha_3 &=16 A^6 (10 p (2 \gamma-1)+9 \mu_m +6 \gamma  (16 \gamma-9) \mu_m ), \notag \\
\alpha_2 &=60 A^7 \left(4 \gamma ^2-1\right) \mu_m,  \notag \\
\alpha_1 &=-40 A^8 (1+2 \gamma ) (4 p+3 (6 \gamma-1) \mu_m ), \notag \\
\alpha_0 &=135 A^9 (1+2 \gamma )^2 \mu_m.  \notag
\end{align}
In contrast to \cite{faulkner71}, here the constant $C_1$ depends on the boundary conditions via the presence of $p$ and $\bar{a}$. \cite{faulkner71} does \textit{not} impose boundary conditions in the form
\eqref{new_boundary_conditions_nearlyincomp} working instead with rather non-physical boundary conditions for the slightly compressible part of the deformation.

\bibliographystyle{model2-names}
\bibliography{bibliography}

\end{document}